\documentclass[11pt,a4paper,twoside]{article}
\usepackage[dvips]{graphics,graphicx,color}
\usepackage{a4,indentfirst,latexsym,amsfonts,amsmath,amssymb,array,subfigure}
\usepackage{wrapfig,delarray,eucal,graphpap}
\usepackage{tabularx,lscape,xr}

\title{Transient thermal effects in solid noble gases as materials for the detection of Dark Matter}

\author{Ionel Lazanu\\University of Bucharest, Faculty of Physics, POBox MG-11\\Bucharest-Magurele, Romania\\ ionel.lazanu@g.unibuc.ro \and Sorina Lazanu\\ National Institute of Materials Physics, POBox MG-7\\ Bucharest-Magurele, Romania\\lazanu@infim.ro}

\begin{document}
\maketitle

\begin{abstract}
Noble solid gases are promising detector materials to be used in the search for dark matter.
In the present paper a systematic analysis of the transient phenomena associated with the stopping of recoils in noble gases in the solid phase is performed for the first time. The investigated energy range of the recoils corresponds to the elastic scattering of WIMPs from the galactic halo in these materials. A thermal spike model, previously developed by the authors, is extended and applied to solid noble gases.  Ionization, scintillation and nuclear energy loss processes are considered and included in the model, as well as the coupling between the subsystems.
The development of the temperature pulse in space and time in solid Ar, Kr and Xe is analysed for different energies of the WIMP, and for different initial temperatures of the material. Phase transitions are possible in particular cases.
The results of the model could be used as supplementary information in respect to ionization and scintillation, for detection and particle identification.
\end{abstract}

\section{Introduction}
In the last decades, the progress in experimental nuclear, particle or astroparticle physics, astronomy, as well as the development of applications in other fields as medicine, space, industry, imposed the development of new methods of radiation detection and measurement.

Noble gases combine several properties that make them very attractive as detection media: good scintillation characteristics, transparency to their own emitted light, high ionisation yield. In the solid phase they are dense, relatively inexpensive and able to be produced as large homogeneous detectors. Being chemical elements with complete electronic shells, they are the simplest solids, have face centered cubic structure and the smallest binding energy between the atoms in the crystalline structure. They have relatively high density in the vicinity of the triple point and are mechanically very soft.

From these elements, xenon presents the advantage of not having long lived radio isotopes, so the radioactive background is very low, and combined with its high value atomic number, self-shielding could be obtained.

The use of noble liquid gases to detect dark matter (DM), including also the direct detection of Weakly Interacting Massive Particles (WIMPs) is currently the subject of intense R\&D carried out by a number of groups worldwide. Argon is much cheaper than other noble gases, and sizeable experience in the handling of massive liquid Ar detectors has been acquired within the ICARUS program \cite{ICARUS}. New experiments, like WARP \cite{WARP}, ArDM \cite{ArDM}, XENON \cite{XENON} Collaborations, DARWIN Consortium \cite{DARWIN}, DM experiments proposed at Dusel facility \cite{DARWIN2}, as well as small collaborations as, e.g. \cite{Bougamont}, are under development considering different technologies and targets. Giant facilities are also proposed: \cite{FAC1, FAC2, FAC3}.

The investigation of the properties of noble gases in the solid phase, in the aim of using them as materials for detectors, has a long history: \cite{Bald 1962, Venables 1966, Venables 1972, Kramer 1972, Niebel 1974, Klein 1976, Kramer 1976, Bolozdynya 1977, Himi 1982, Kubota 1982, Aprile 1994}; in spite of the promising results, the subject was nearly abandoned for a period of time.

Recently, the possibility to use solid noble gases for DM experiments was reconsidered - see for example the talk of Yoo, related to the possible use of xenon \cite{Yoo}, in particular for WIMPs and solar axions, or for processes induced by neutrinos (neutrinoless double beta decay,  neutrino coherent scattering from solar reactions or supernova, for example). Due to the crystalline structure of solid noble gases, the possible channelling induced by WIMPs could also be important; its contribution was recently calculated \cite{Bozorgnia 1011.6006} in a similar manner as for usual crystals \cite{Bozorgnia  JCAP 1011:019, Bozorgnia JCAP 1011:028, Bozorgnia  JCAP 1011:029}.

In the present paper we model the transient thermal effects induced by the energy deposited by projectiles in the solid noble gases Ar, Xe and Kr. The projectiles are recoils produced by the elastic scattering of WIMPs in these materials, and the process is related to their direct detection. After a short discussion on kinematic aspects of the interactions of WIMPs with the nuclei of the targets, the transient processes induced by the recoils resulting from WIMPs interactions are quantitatively analysed in the frame of a thermal spike model developed previously by the authors \cite{nim 2010}, and modified to consider the peculiarities of solid noble gases. In the model, ionization and nuclear energy losses, as well as scintillation are taken into account. The  atomic and electronic subsystems are coupled through the electron - phonon interaction.  The time and space dependencies of the lattice and electron temperatures near the recoil trajectory are calculated for each material.
The possibility of producing a local phase transition is also evidenced by the results. A comparison with the transient effects produced by the same WIMPs in Si and Ge is performed. Possible consequences for applications are considered.

\section{Some practical aspects of WIMP's searches using direct detection experiments}
The existence of DM is one of the possible indications of the physics beyond the Standard Model with WIMPs as heavy components, with mass in the range of tens GeVc$^{-2}$ up to TeVc$^{-2}$. The challenge is to discover WIMPs through indirect detection of their annihilation products, direct detection of their scattering with nuclei, and/or production at high energy accelerators.

M. Perl gave some convincible arguments supporting the idea of direct detection for DM \cite{Perl 2009}. If the WIMPs represent the dominant contribution to DM, their density in the galactic halo is estimated to be about 0.3 GeVc$^{-2}$m$^{-3}$ in our neighbourhood and their velocity should be in the range of 200 km/s, hence they are dominantly nonrelativistic. A detailed model of WIMP interaction with nuclei was developed by Lewin and Smith \cite{Lewin 1996}.

The WIMPs from the halo elastically scattered on target nuclei of a terrestrial detector leave a small amount of energy (less than 100 keV) to the recoil nucleus.

The recoil energy dependence on the WIMP mass, considering only the elastic scattering, for different targets, is presented in Figure 1. In all the curves, the velocity of the WIMP is taken 280 km/s, and an average of the recoil energies over scattering angles is considered.

\begin{figure}[!htb]
\centering
\includegraphics[width=0.6\textwidth]{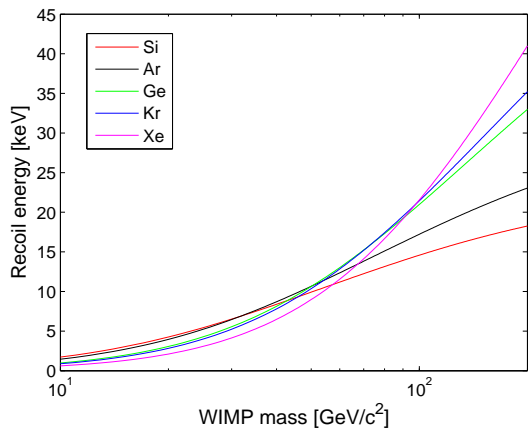}
\begin{small}
\caption{Recoil energy as a function of the WIMPs mass, for elastic scattering. Target materials are Ar, Kr, Xe, Si and Ge.}
\label{fig1}
\end{small}
\end{figure}

The range of recoil energies in all the targets considered is very narrow for WIMP masses up to 60 GeV/c$^2$, and in all cases is up to tens of keV. Recently, the CDMS II Collaboration reported the possible existence of two candidate events for WIMPs at recoil energies of 12.3 keV and 15.5 keV respectively, that passed all selection criteria \cite{CDMS}.

\section{Transient processes induced by a projectile and their modelling }
The interaction processes of different particles in crystalline solids depend both on projectile (mass, electrical charge, kinetic energy) and target characteristics (density, atomic number, and bonding energy). Their effect is disorder production, either transient or permanent. Nuclear stopping dominates the energy loss at low kinetic energy of the incident ion or at the end of range of the recoil produced in particle - atomic collision, while at higher energies electronic stopping becomes increasingly important.
The atom recoiling from the interaction with the incident particle generates a cascade of atomic collisions. In spreading the primary recoil energy over a small interaction region, the cascade presumably produces a local heating effect.

Such phenomena have first been considered by Seitz \cite{Seitz 1949, Seitz 1956} and by Brinkman \cite{Brinkman 1954}, that developed a theoretical formalism for the regime of high nuclear energy loss. In the nineties, it has been shown that the energy transferred by an energetic incoming particle to electrons could produce lattice defects \cite{Toulemonde 1993, Szenes 1995, Vetter 1998}. The regime of comparable electronic and nuclear energy losses was studied by the authors considering the distinct contributions from electronic and nuclear sources in semiconductors \cite{nim 2010}, and also the case of dominant nuclear contribution to heat \cite{nim 2011}. This last regime corresponds to very low energy for the projectile or recoils and appears at the end of their range. In crystals, in this case the mean free path between collisions is comparable or lower than the lattice constant.

The energy lost by the incoming particle is imparted between the electronic and lattice (nuclear) subsystems of the target. After the processes by which the recoil (or projectile) loses its energy in the medium, the two subsystems have different temperatures and are coupled through a term that is a measure of the energy exchange, the electron - phonon coupling. Details on the model describing the time and space development of the heated regions in the two subsystems, of its general hypotheses, as well as the influence of physical characteristics of the target are discussed in the previous papers \cite{nim 2010} and \cite{nim 2011}. The processes are studied in a thin layer, perpendicular to the track, and a cylindrical symmetry is considered - see in Ref. \cite{nim 2011} a discussion on the applicability of the cylindrical and spherical spike. The localized regions of the medium characterized by departure from equilibrium due to the energy transfer from the projectile toward electrons and nuclei respectively are generally different because the mechanisms of interaction and the kinematics are distinct. In the application of the model to noble solid gases where a WIMP interacts with a nucleus of the target, the possibility that local phase transitions occur is taken into account.

The temperatures of the electronic $T_e$ and atomic $T_a$ subsystems as functions on the distance to the projectile's track, and time after its passage, are solutions of two coupled partial differential equations:

\begin{equation}
\begin{split}
 C_e \left( {T_e } \right)\frac{{\partial T_e }}{{\partial t}} = \frac{1}{r}\frac{\partial }{{\partial r}}\left[ {rK_e \left( {T_e } \right)\frac{{\partial T_e }}{{\partial r}}} \right] - g\left( {T_e^p  - T_a^p } \right) + A\left( {r,t} \right) \\
 C_a \left( {T_a } \right)\frac{{\partial T_a }}{{\partial t}} = \frac{1}{r}\frac{\partial }{{\partial r}}\left[ {rK_e \left( {T_e } \right)\frac{{\partial T_a }}{{\partial r}}} \right] - g\left( {T_a^p  - T_e^p } \right) + B\left( {r,t} \right) \\
 \end{split}
 \end{equation}
where $C_e$, $C_a$ are volumetric heat capacities of electronic and atomic systems and $K_e$, $K_a$ are their corresponding thermal diffusivities. The two sub-systems are coupled through the term: $g\left( T_a^p  - T_e^p  \right)$, with $p$ around 1 at RT (Newton's law of cooling) and $p$ generally higher at very low temperatures (e.g. $p$ = 5 $-$ 6 in Si \cite{nim 2010}). The source terms satisfy the conservation laws:

\begin{equation}
\begin{split}
\int^\infty_0 dt\int^\infty_0 2\pi r A(r,t)dr&=S_e(1-f)\\
\int^\infty_0 dt\int^\infty_0 2\pi r B(r,t)dr&=S_n
\end{split}
\label{eq1}
\end{equation}
with $S_e$, $S_n$ the electronic and nuclear stopping powers respectively, evaluated with SRIM, and $f$ the fraction of the ionization energy loss used in the scintillation - see the discussion below. The absorption of energy during local phase transitions is kept into account.
The characteristics of the target as thermal conductivity, specific heat and electron-phonon coupling are considered in the whole range of temperatures of interest, using published data, for each of the solid noble gases.

Details on the model describing the time and space development of the heated regions in the two subsystems, of its general hypotheses, as well as the influence of physical characteristics of the target are discussed in the previous papers \cite{nim 2010} and \cite{nim 2011}. An hypothesis of the model is that the electronic and atomic sources, related to the initial energy distribution in these systems, $A(r,t)$ and $B(r,t)$ respectively, have similar analytical expressions.

The noble gases are also scintillating substances. The energy transferred to electrons is used in two ways: as excitation and ionization. In the excitation process, an electron is raised  to a higher energy state; it subsequently returns to its original state via a cascade process resulting in the emission of photons having discrete and characteristic energies. The secondary electrons from the ionization process may, in their turn, generate new pairs or excitations. The detailed mechanism is described in different papers, see for example the book of Aprile and co-workers \cite{Aprile 2006}. The excitation energy released during these processes is manifested in the emission of a visible - UV radiation, or through the production of heat. The contribution of the scintillation is also included in the present calculations, by considering the electronic energy loss as a sum of energy converted into scintillation and into ionization.

\section{Numerical results and discussions}
The atoms of rare gas elements have completely filled outer electronic shells. In crystalline form, the cohesive forces between the atoms are the weak van der Waals forces. The valence electrons are tightly bound, and consequently these noble gas solids are insulators.

Because there are no free electrons in insulators, the knowledge of the values of the parameters of the electronic sub-system is difficult and no experimental values exist in the literature. Following the reasoning of Toulemonde \textit{et al.} \cite{Toulemonde 2006}, in the present paper we assumed the electron heat capacity $C_e$ = 1 Jcm$^{-3}$K$^{-1}$, and the thermal conductivity $K_e$ = 2 Wcm$^{-1}$K$^{-1}$, both temperature independent. The electron-phonon coupling constant $g$ is related to the mean free path of electrons, and this last has been reported in Ref.  \cite{Toulemonde 2006} to have values in the range 1 $-$ 5 nm. The corresponding coupling constant, in the case of linear heat transfer between the two sub-systems, is eventually derived to be $g = 8 \times 10^{12}$ Wcm$^{-3}$K$^{-1}$, the same for all noble solid gases. $g$ was supposed to be temperature independent, and the law of heat transfer linear, i.e. $p = 1$ in formula (1).

The temperature dependence of the values of the thermal parameters characterising the atomic sub-system were fitted using the experimental data reported in the literature for each of the noble gas solids of interest.

For solid Ar, the data on density are from the papers \cite{Dobbs 1957} and \cite{Dobbs 1958}, the data on specific heat are from \cite{Dobbs 1957} and \cite{Finegold 1969}, and for the thermal conductivity the results reported in \cite{White 1956, White 1958, Clayton, Christen 1975} were used.

The curve fits used in the present paper are presented in Figure 2 together with the experimental data for the lattice specific heat and thermal conductivity in the whole range of temperatures of interest.

\begin{figure}[!htb]
\centering
\subfigure{
\includegraphics[width=0.47\textwidth]{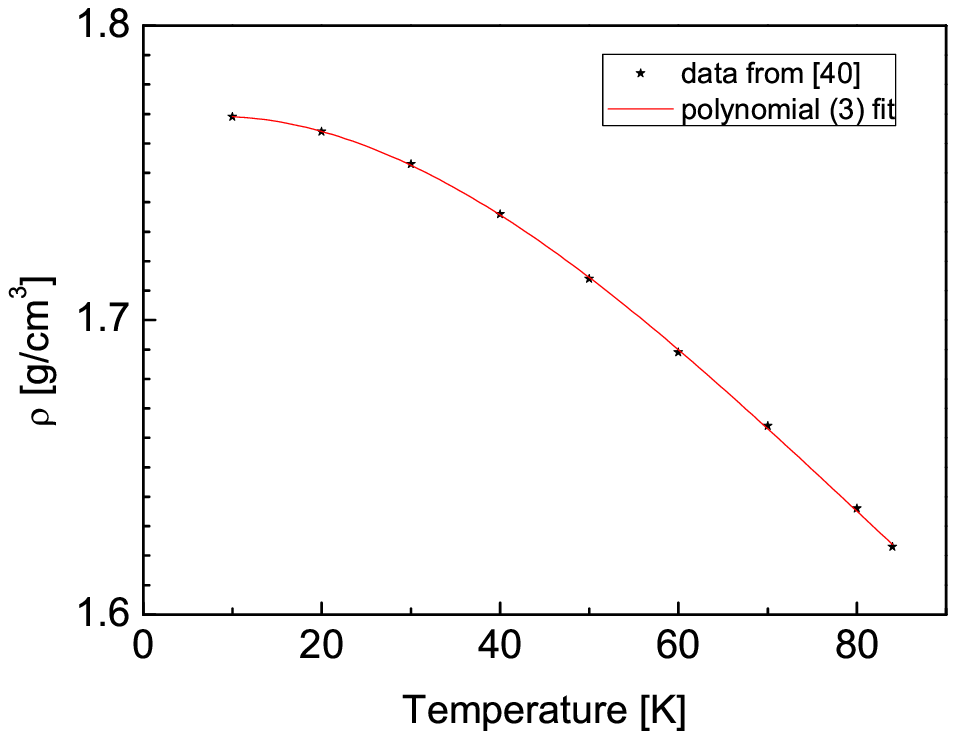}
\label{fig:subfig2a}
}
\subfigure{
\includegraphics[width=0.47\textwidth]{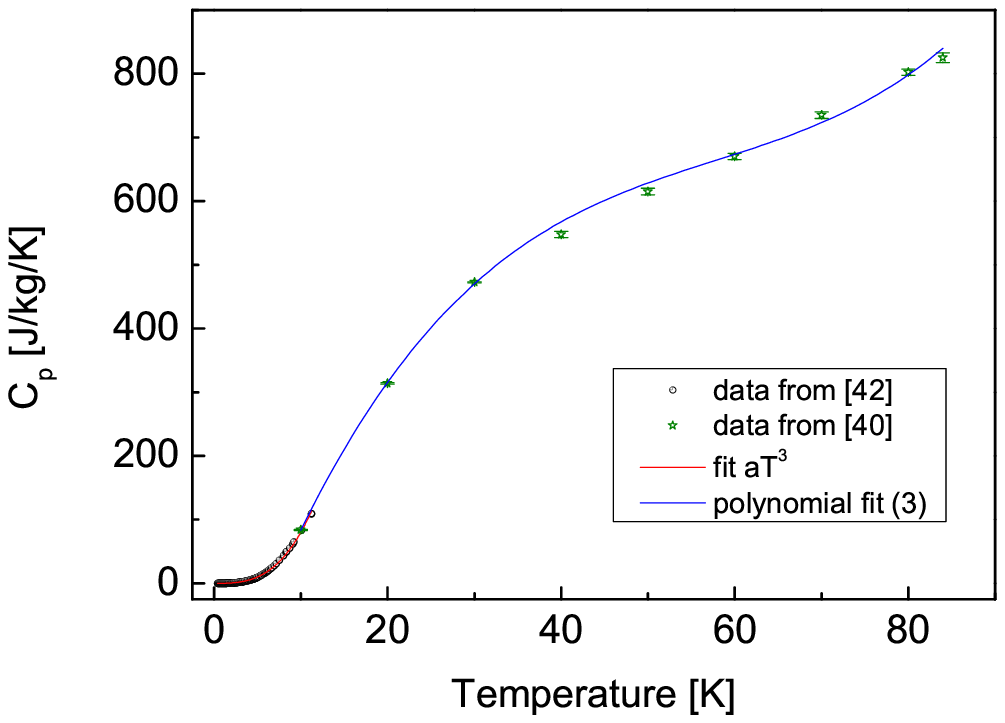}
\label{fig:subfig2b}
}
\subfigure{
\includegraphics[width=0.47\textwidth]{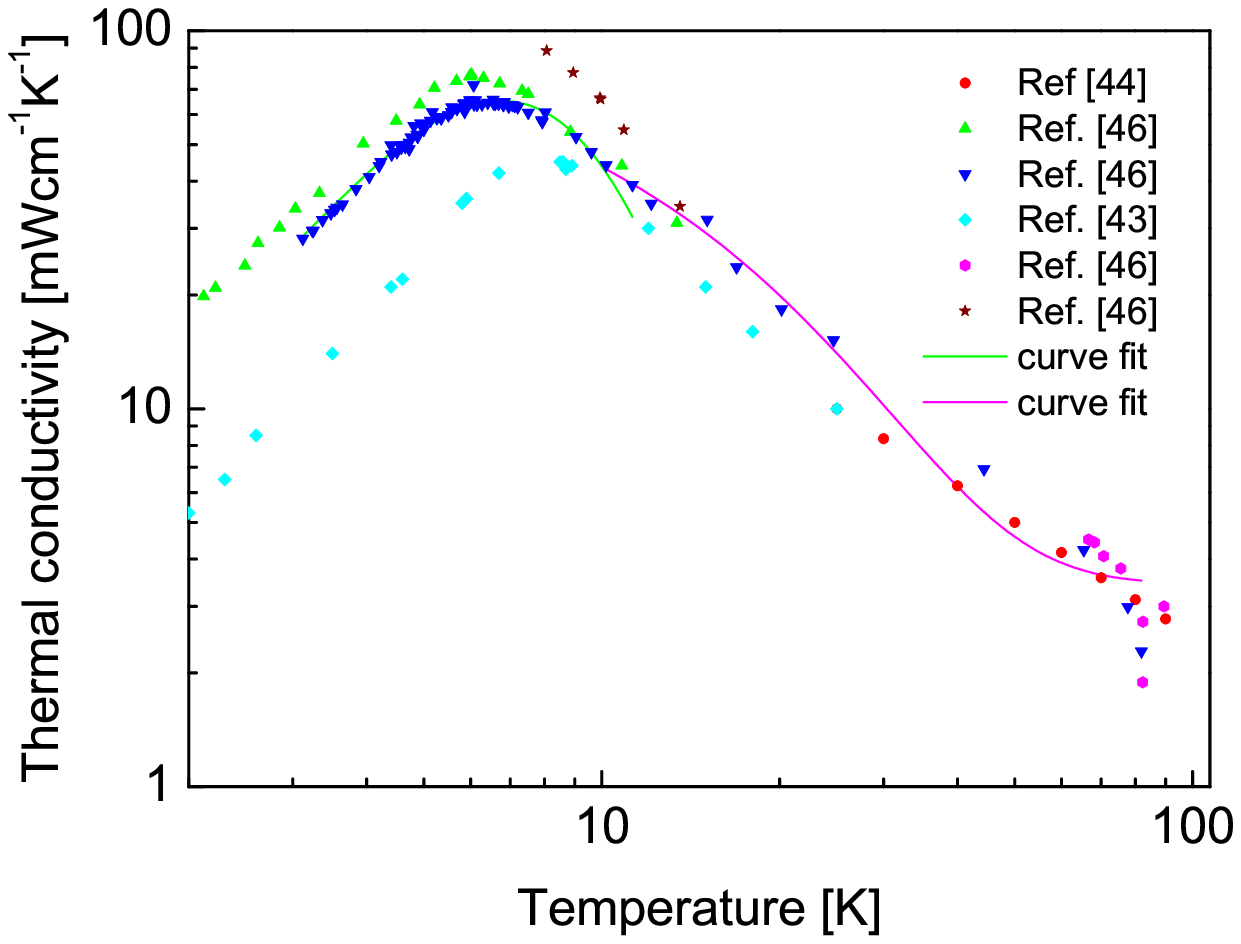}
\label{fig:subfig2c}
}
\begin{small}
\caption{Temperature dependence of the density, specific heat and thermal conductivity of solid Ar.}
\label{fig2}
\end{small}
\end{figure}

For solid Xe, we used data on resistivity from Refs. \cite{Aprile 2006, Eatwell 1961, Sear 1962}, data on specific heat from \cite{Fenchel 1966} and  \cite{Trefny 1969} and for thermal conductivity, the data from \cite{Purskii 2004} respectively. The experimental data on density, lattice specific heat and thermal conductivity, together with the curve fits, are presented in Figure 3. For solid Xe we did not find experimental data on thermal conductivity at very low temperatures. It is known \cite{Trimm} that for isolating crystals, the maximum of $K_T$ occurs approximately at $T_D/10$, where $T_D$ is the Debye temperature, and that $K_T$ gradually decreases as $T \to\ 0$ K. In the present calculations, the curve fit for $K_T$ was extended beyond the experimental data, as shown in Figure 3c, with dotted line, up to $T_D/10$, and for lower temperatures a linear dependence on $T$ was supposed, represented also dotted.

\begin{figure}[!htb]
\centering
\subfigure{
\includegraphics[width=0.47\textwidth]{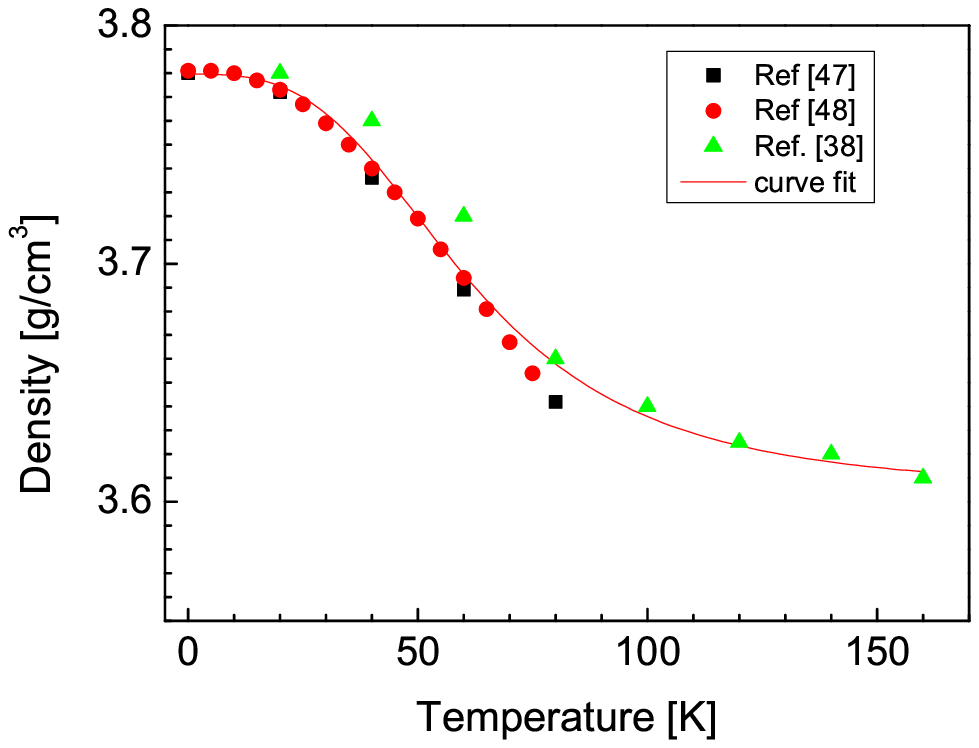}
\label{fig:subfig3a}
}
\subfigure{
\includegraphics[width=0.47\textwidth]{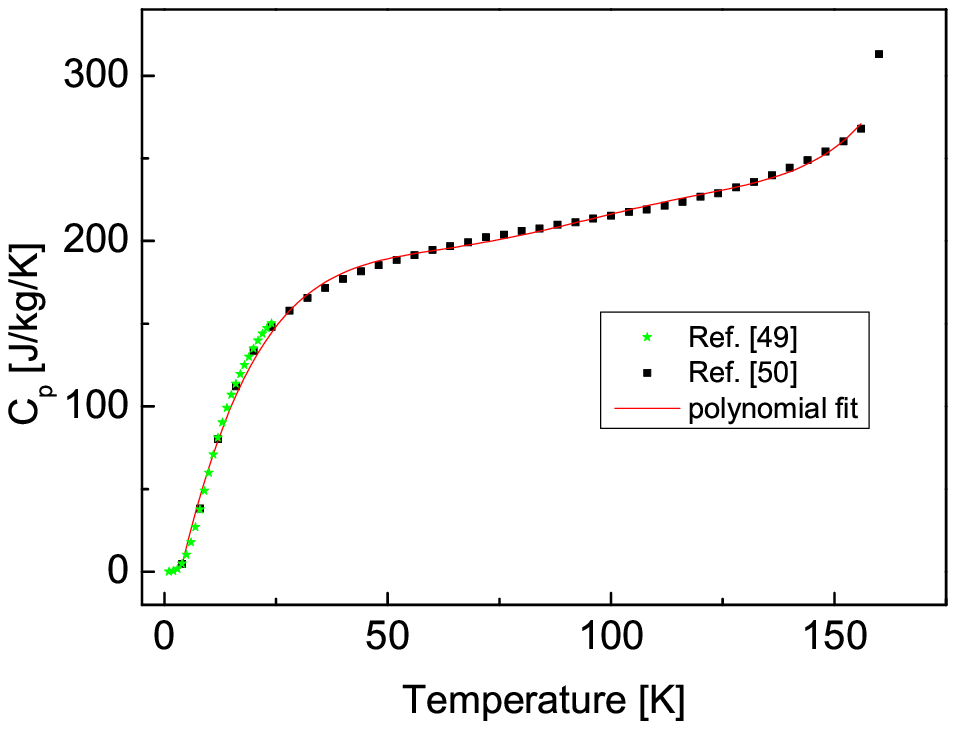}
\label{fig:subfig3b}
}
\subfigure{
\includegraphics[width=0.47\textwidth]{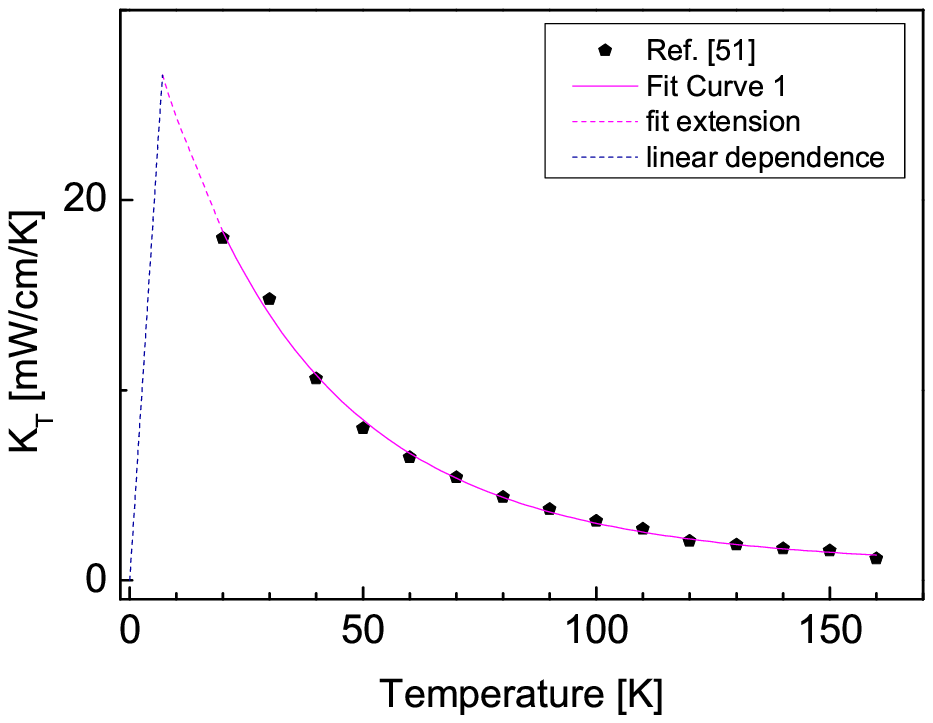}
\label{fig:subfig3c}
}
\begin{small}
\caption{Temperature dependence of the density, specific heat and thermal conductivity of solid Xe.}
\label{fig3}
\end{small}
\end{figure}

For Kr, the density data from \cite{Aprile 2006} and \cite{Figgins 1960} were used; the temperature dependence of the heat capacity was extracted from the works \cite{Finegold 1969} and \cite{Beaumont 1961}. The thermal conductivity data from \cite{White 1958} and \cite{Krichikov} were considered. In the model calculations, the data were smoothed out and interpolated - see Figure 4.

\begin{figure}[!htb]
\centering
\subfigure{
\includegraphics[width=0.47\textwidth]{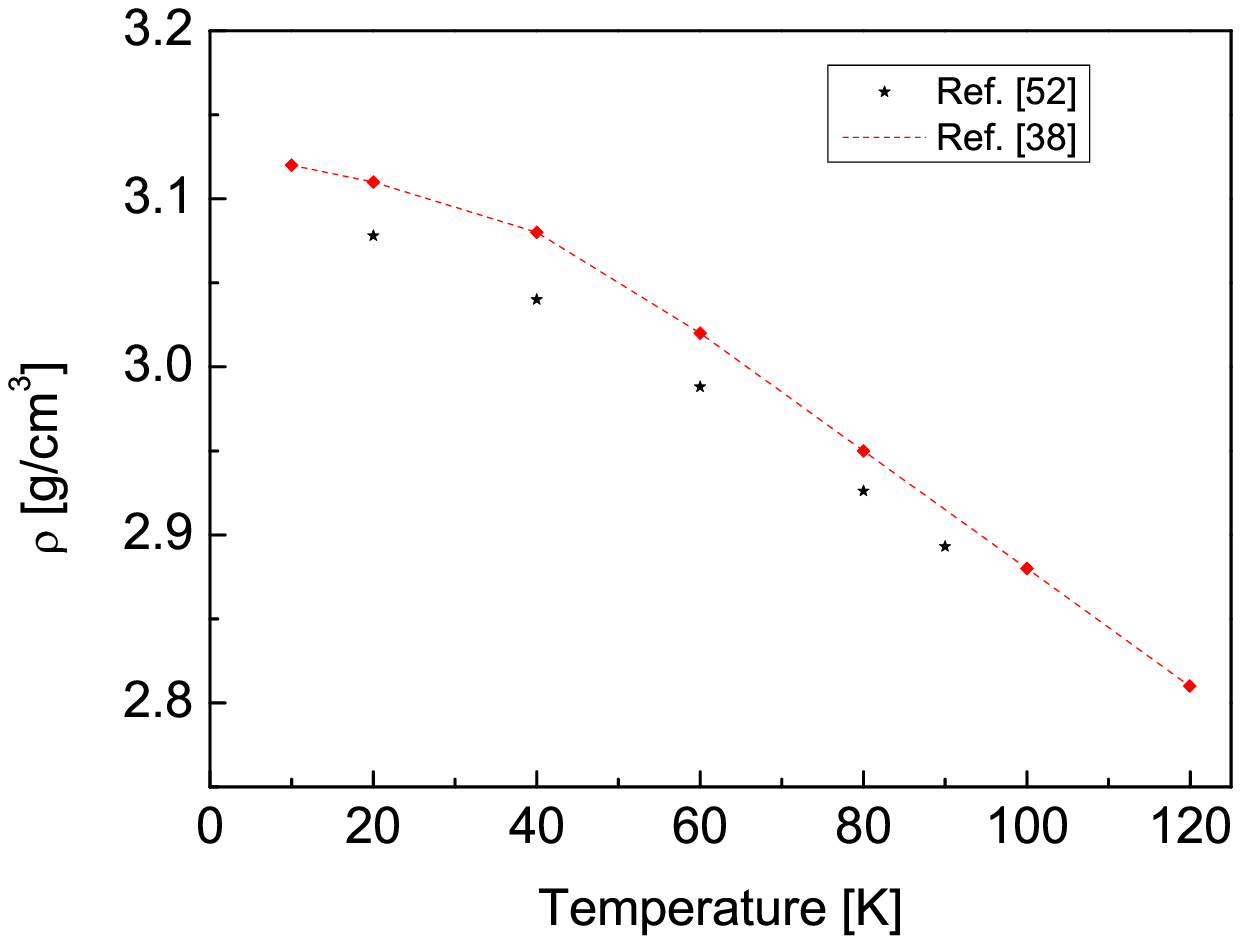}
\label{fig:subfig4a}\
}
\subfigure{
\includegraphics[width=0.47\textwidth]{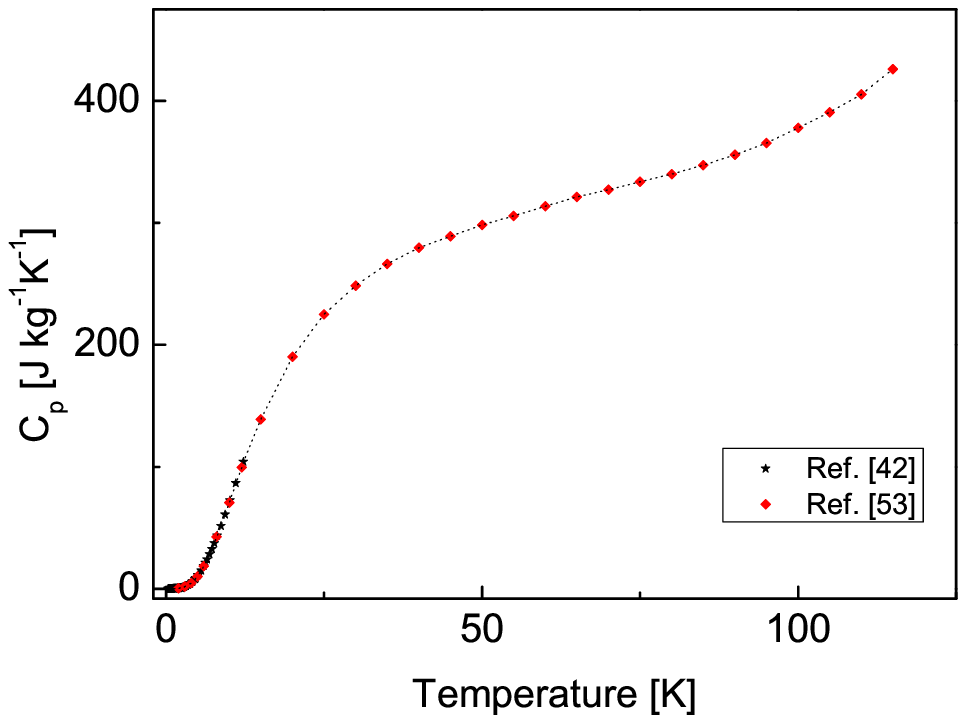}
\label{fig:subfig4b}
}
\subfigure{
\includegraphics[width=0.47\textwidth]{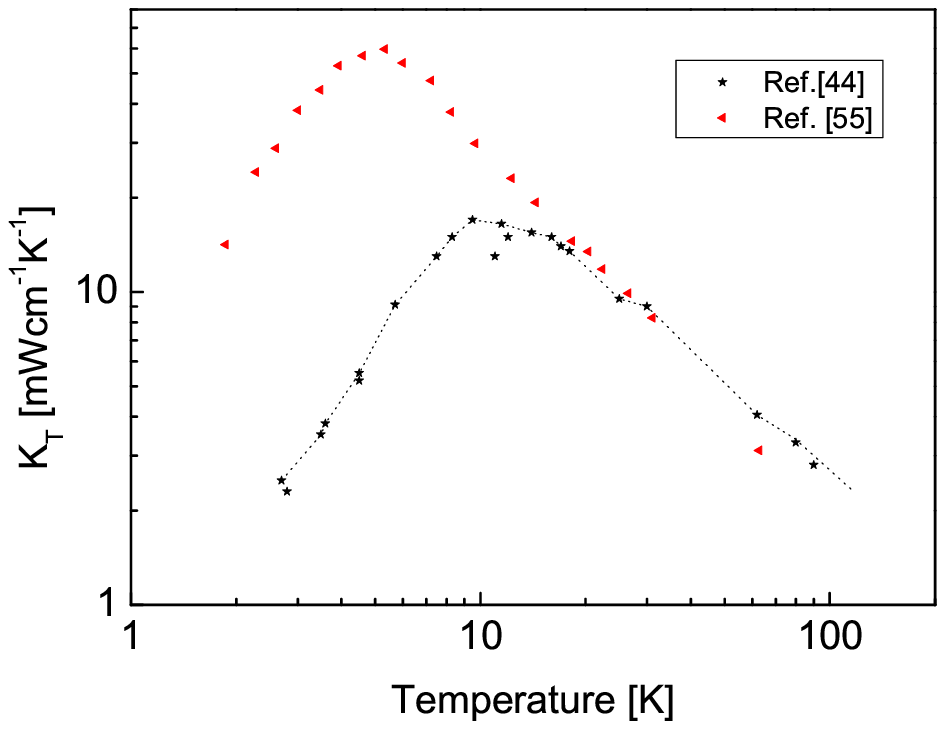}
\label{fig:subfig4c}
}
\begin{small}
\caption{Temperature dependence of the density, specific heat and thermal conductivity of solid Kr.}
\label{fig4}
\end{small}
\end{figure}

The energy transferred to the material in an interaction of an incoming particle could heat up locally the solid up to the melting point, and even further. The data on latent heats for solid - liquid and liquid - gas transitions respectively, listed in Table 1 for Ar, Xe and Kr are from Ref. \cite{Aprile 2006}.

\begin{table}[!htb]
\caption{Characteristics of the phase transitions in Ar, Kr and Xe.}
\label{tab1}

\centering
\begin{tabular}{|c|c|c|c|c|}\hline
   &                &               &               &              \\
   & Melting point & Latent heat of & Boiling point & Latent heat of\\
   & $T_T$[K]      & fusion at $T_T$& at 1 atm      & vaporization  \\
   &               &$\lambda_T$[kJ kg$^{-1}$]& $T_b$[K] & $\lambda_b$ at $T_b$[J kg$^{-1}$]\\
   &                &               &               &              \\
\hline
Ar & 83.80 & 29.44 & 87.26 & 163.20 \\
\hline
Kr & 115.79 & 19.52 & 119.74 & 107.70\\
\hline
Xe & 165.03 & 17.48 & 169.00 & 96.29 \\
\hline
\end{tabular}
\end{table}


In the modelling, the fact that part of the energy transferred to the electronic system is used for scintillation was also kept into account: in Ar and Xe the data from Refs. \cite{Ar, Xe} respectively were used. No experimental data on the percentage of the electronic energy deposited by projectiles which is converted into scintillation exist, to our knowledge, in the literature for Kr, and an average of 40 \% of the ionising energy loss was supposed to contribute to the transient phenomena analysed, independent on the recoil energy.

The response of the crystal to the passage of a particle depends both on the characteristics of the material and on the energy loss of the projectile. The range of selfrecoils in solid noble gases, for the energies of interest here, is between tens and hundreds nm, hence hundreds of lattice constants, so that the use of the cylindrical thermal spike model is justified. In Figures 5 and 6, the development of the heated regions, as space and time dependencies, is illustrated for a primary self recoil of 20 keV, which resulted from the interaction with the WIMP, both in the atomic and electronic subsystems, in Ar and Xe respectively. In both cases, the ambient temperature is 2 K.

\begin{figure}[!htb]
\centering
\subfigure{
\includegraphics[width=0.47\textwidth]{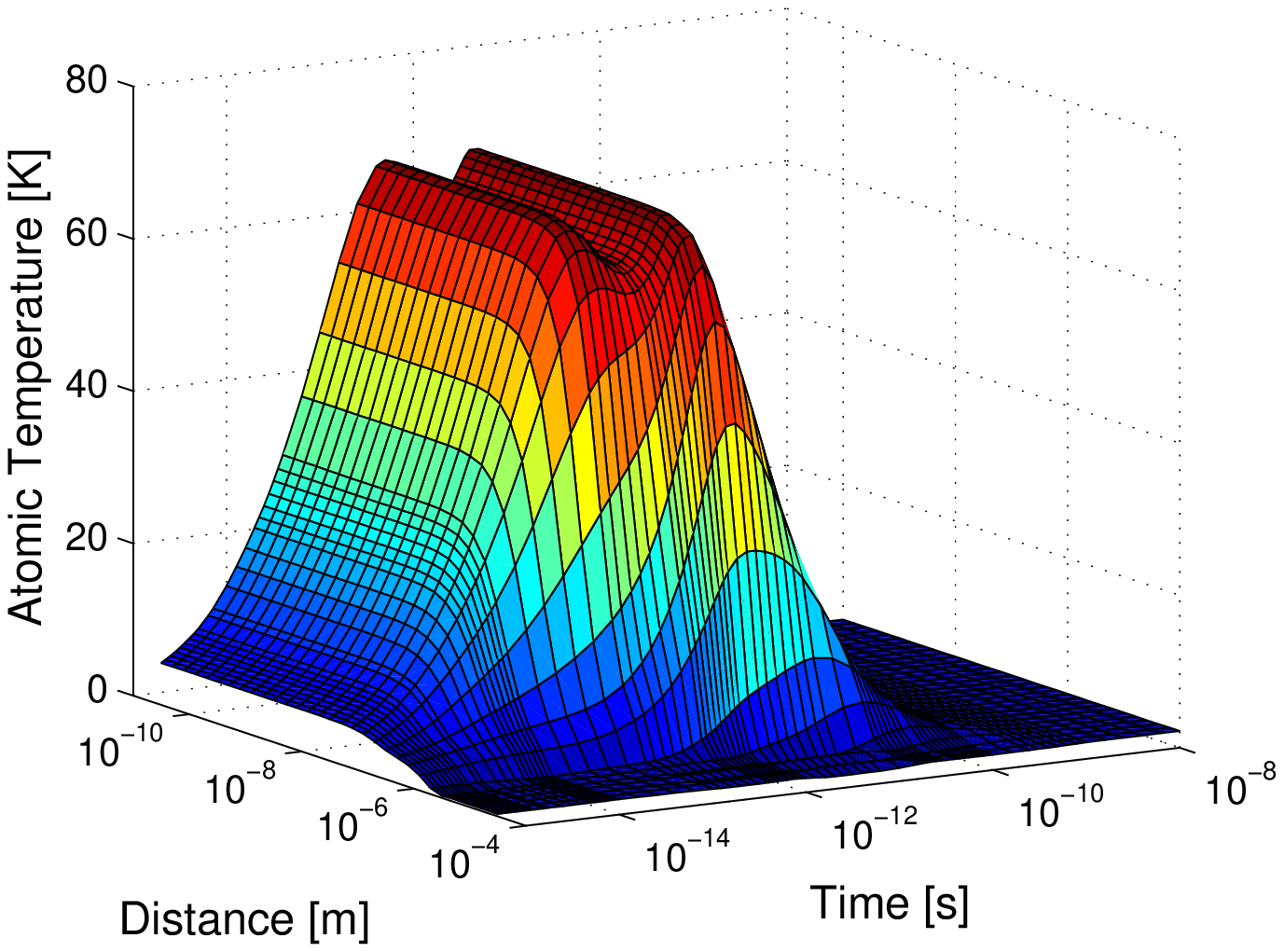}
\label{fig:subfig5a}
}
\subfigure{
\includegraphics[width=0.47\textwidth]{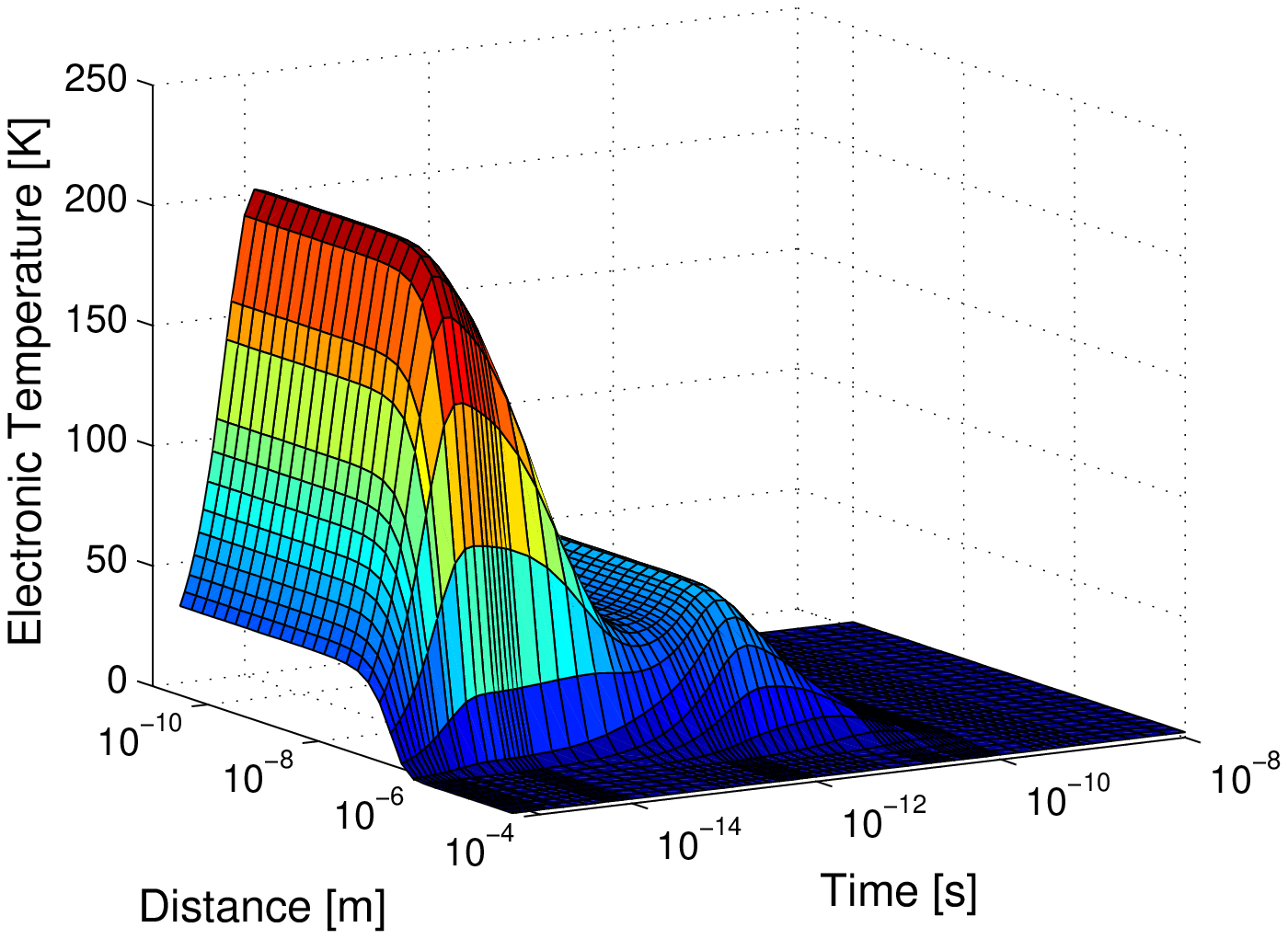}
\label{fig:subfig5b}
}
\begin{small}
\caption{Time and space development of the heated regions in the atomic (left) and electronic (right) subsystems of Ar at 2 K, produced by an argon recoil of 20 keV.}
\label{fig5}
\end{small}
\end{figure}

Due to the electron - phonon coupling, the curves present a structure with two peaks (or one peak and a shoulder). The time developments of the heat pulses in the two subsystems are in agreement with the time scales of the characteristic processes \cite {Itoh 2009}. In the range of recoil energies of interest in the present paper, in Ar, where the ionization and nuclear energy loss are not so different (in the ratio 0.1 $-$ 0.2), the two peaks for the atomic temperature have comparable heights, while for Xe, where the ratio $(dE/dx)_{ioniz}/(dE/dx)_{nucl}$ is around 0.02, only one peak is visible in the atomic temperature.

\begin{figure}[!htb]
\centering
\subfigure{
\includegraphics[width=0.47\textwidth]{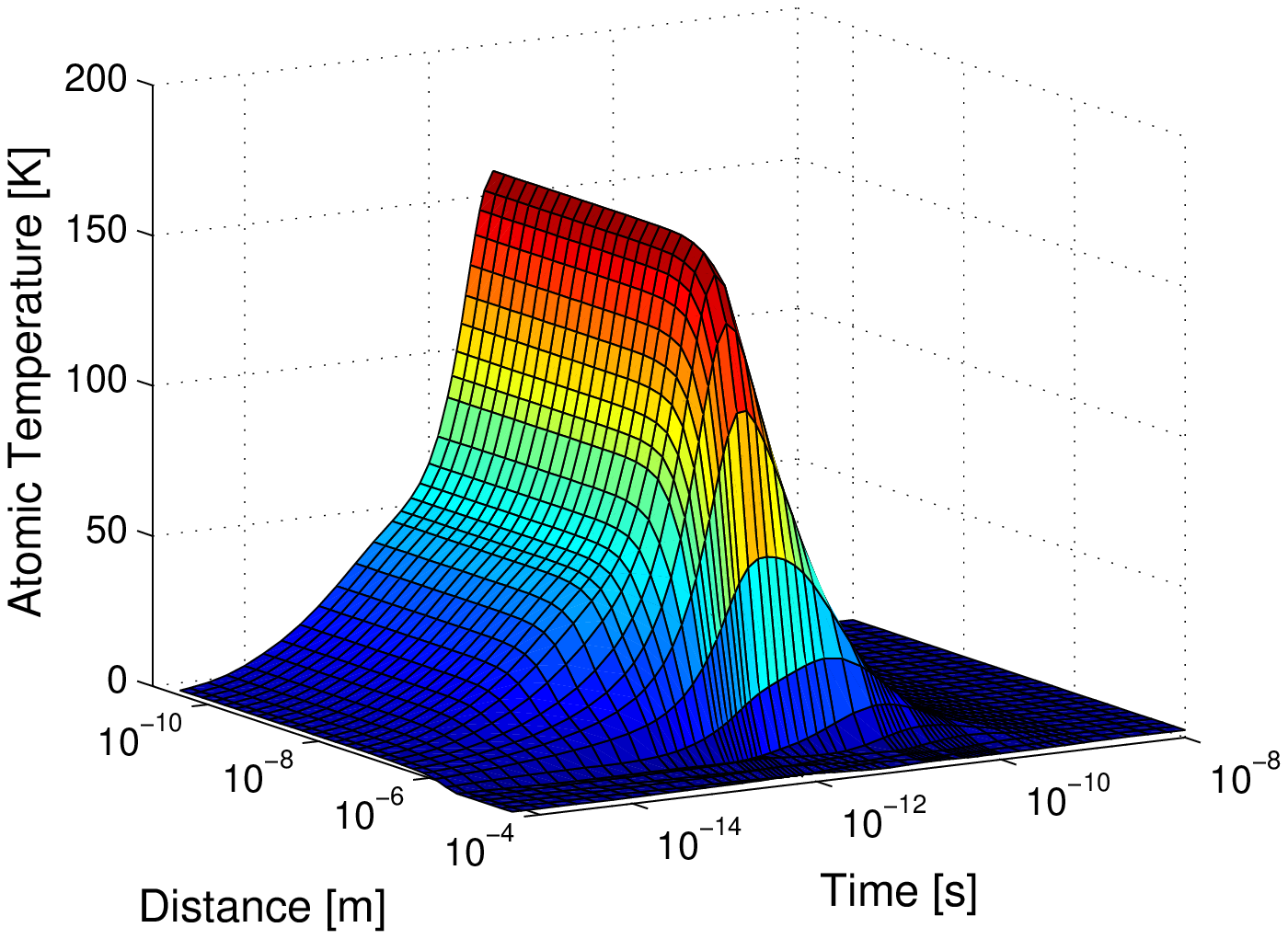}
\label{fig:subfig6a}
}
\subfigure{
\includegraphics[width=0.47\textwidth]{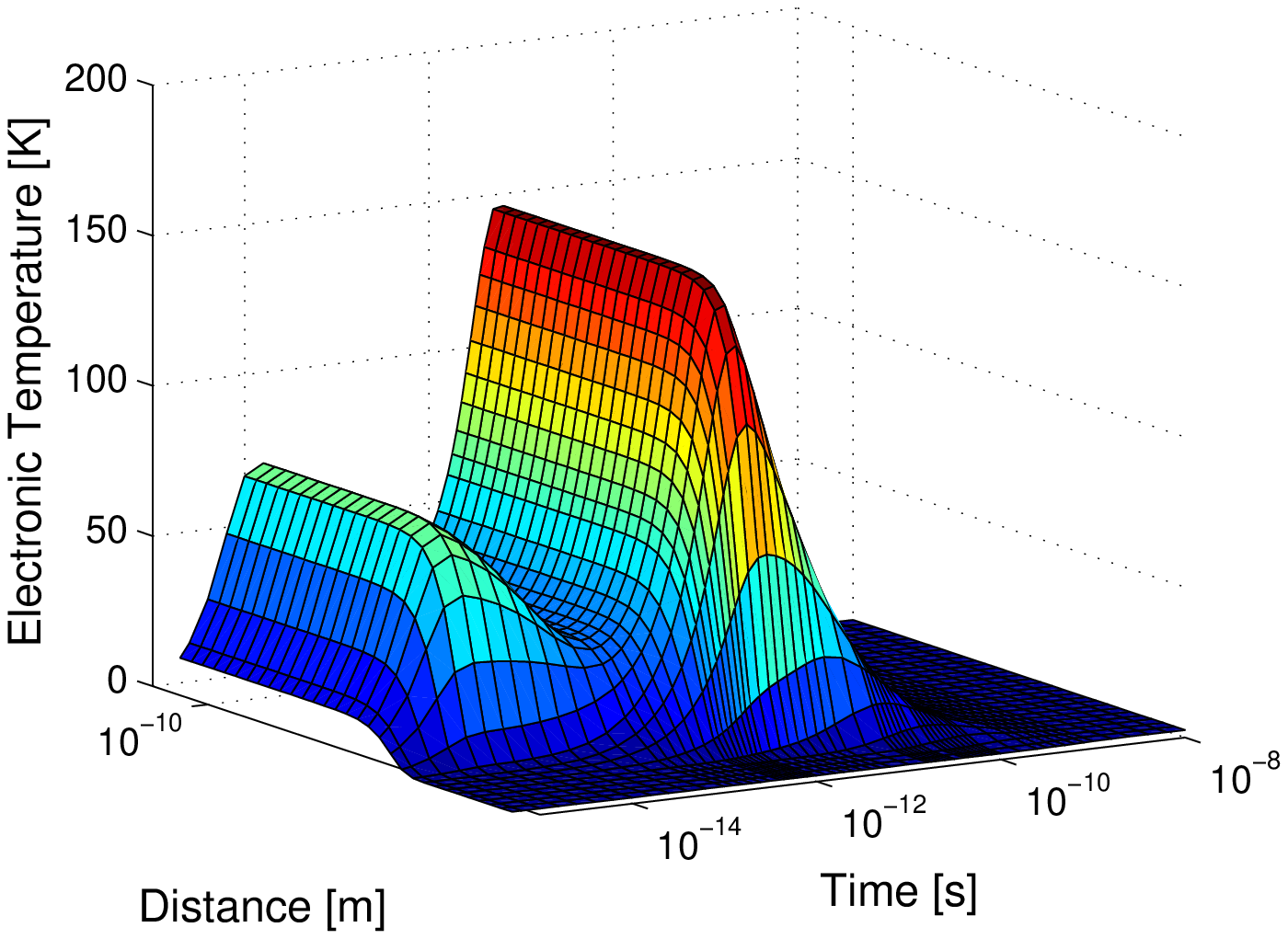}
\label{fig:subfig6b}
}
\begin{small}
\caption{Time and space development of the heated regions in the atomic (left) and electronic (right) subsystems of Xe at 2 K, produced by a recoil of 20 keV.}
\label{fig6}
\end{small}
\end{figure}

The scintillation is delayed in respect to the thermal spike; it is included in the energetic balance but is not observable in these distributions. The dynamics of these phenomena is different in an applied electric field or during mechanical compression, but this problem is not discussed here.

An important point is to put in evidence the capability of the model to discriminate between recoils generated by different particles in the energy range considered here and also between recoils and other projectiles that interact directly. If the nonionizing energy loss (NIEL) is used as a quantitative average measure of the rate of energy loss due to atomic displacements, thus, a useful correlation between different particles could be done. For details see the original formalism developed by Lindhard and co-workers \cite {Lindhard}  as well as subsequent contributions \cite{Summers, laz}.

In figure 7 the energy dependence (in the range of interest for WIMPs detection) of NIEL in solid Ar, Kr, and Xe is presented both for self recoils and protons.  The lattice constants of the solid noble gases are from Ref. \cite{chem}. The calculations were performed in the frame of an analytical model \cite{messenger}, based on Thomas-Fermi potentials, and which was proved to have a large validity \cite{Como noi 2007}. As shown in Ref. \cite{jun}, the NIEL induced by electrons in different materials between carbon and mercury is an increasing function of energy, which saturates at about $2 \times 10^{-4}$ MeVcm$^2$/g for an energy range up to 1 GeV. The NIEL produced by photons is lower than the electron one \cite{Ginneken}.  The energy dependencies of NIEL for electrons and photons are not included in the figure.

\begin{figure}[!htb]
\centering
\includegraphics[width=0.6\textwidth]{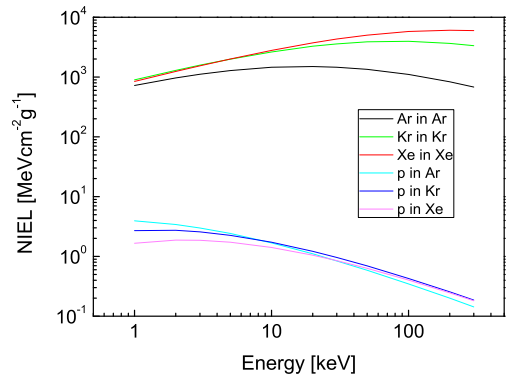}
\begin{small}
\caption{Energy dependence of the NIEL of selfrecoils and protons in solid noble gases Ar, Kr and Xe.}
\label{fig7}
\end{small}
\end{figure}

Another aspect considered here is the difference in the thermal effects produced by electron interactions and nuclear-recoil interactions with the same energy. In Fig. 8, the space and time dependence of the temperature of the heated region, produced by an electron of 20 keV in solid Ar of initial temperature 2K is presented, both in the atomic and electronic systems. The electronic stopping power is from the ESTAR programme \cite{ESTAR}. As could be seen, electrons of the same energy as argon self-recoils produce a negligible thermal effect. For comparison, see also Figure 5.

\begin{figure}[!htb]
\centering
\subfigure{
\includegraphics[width=0.47\textwidth]{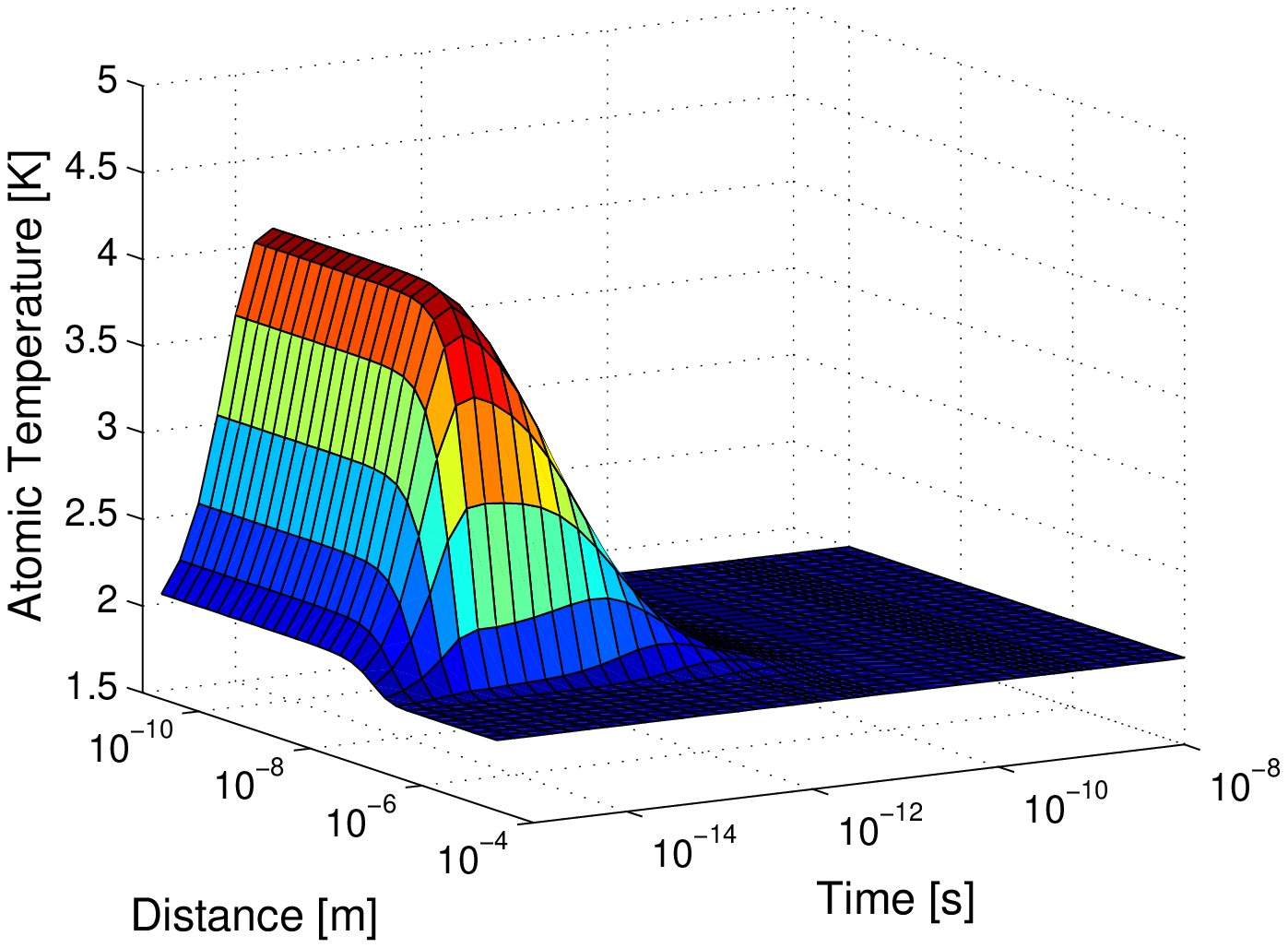}
\label{fig:subfig8a}
}
\subfigure{
\includegraphics[width=0.47\textwidth]{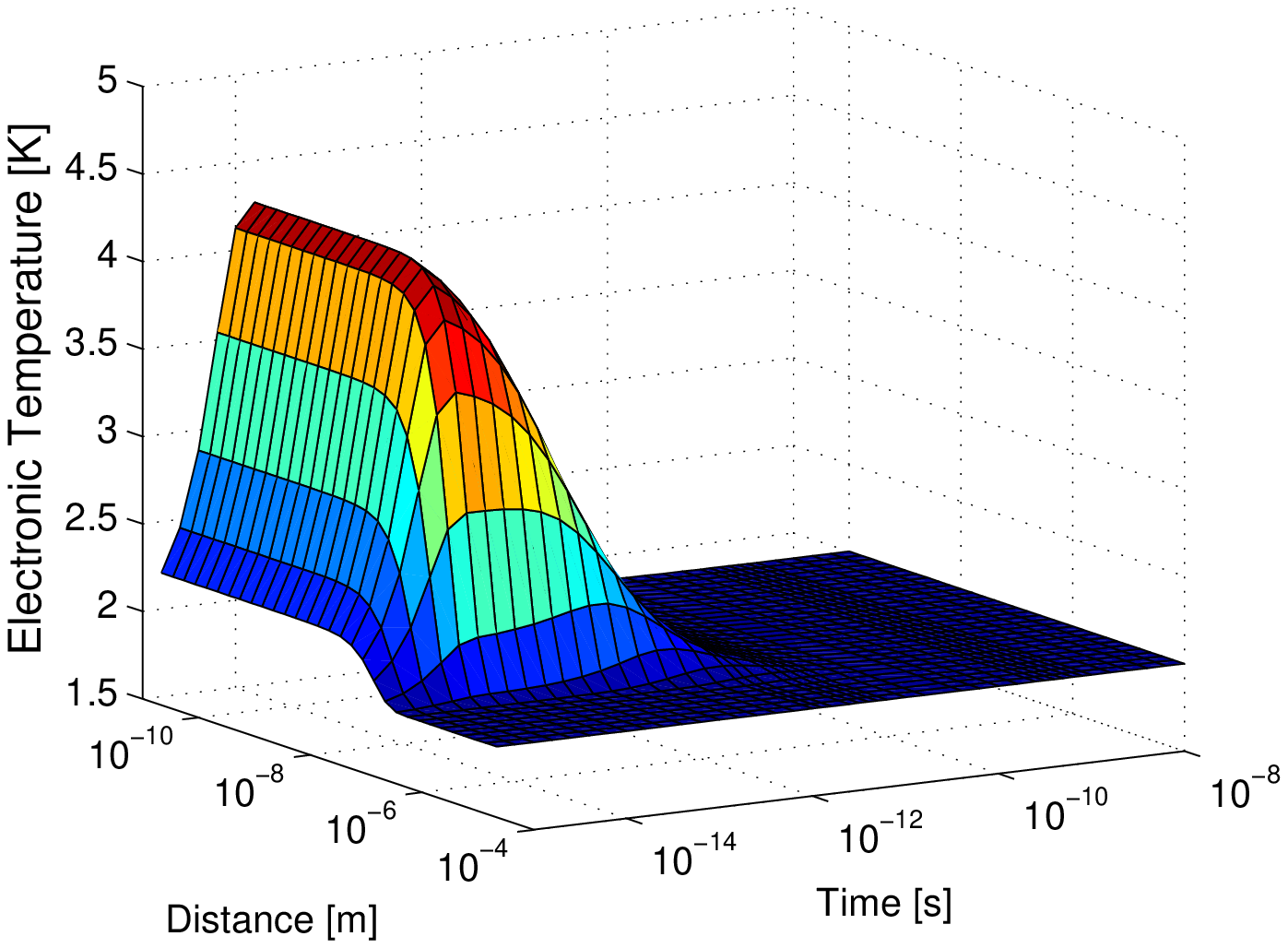}
\label{fig:subfig8b}
}
\begin{small}
\caption{Time and space dependence of the atomic (left) and electronic (right) temperature produced in solid Ar by a 20 keV electron.}
\label{fig8}
\end{small}
\end{figure}

Although the melting temperature of solid Xe is around 80 K higher than the corresponding one of Ar, the 20 keV recoil in Xe produces a partial local heating of the material until it gets near the melting temperature, in contrast with Ar. We would like also to stress that the phase transition is related to the temperature of the atomic subsystem.

We further studied the effect of a WIMP interaction in a Xe crystal kept at liquid nitrogen temperature. The WIMP of 100 GeV/c$^2$ and 280 km/s is supposed to produce a recoil of 25 keV kinetic energy as a consequence of the elastic scattering. In this situation, from the model calculations we found that the solid partially melts.

The development of the heated region in the solid is obtained as solution of eqs. (1 - 2). If at the melting temperature the energy available in the atomic system surpasses the latent heat of fusion, a quantity of substance is transformed into liquid. If the energy deposited is high enough, the possible occurrence of a partial phase transition into vapours or gas must also be considered. In the present case, the phenomena taking place in the liquid are more complex in respect to the superheated liquids, analysed first by Glaser \cite {Glaser}, and subsequently by Seitz \cite{Seitzbub}. The operation of bubble chambers is based on the phase transition induced by the passage of particles in a volume of superheated liquid, kept at fixed pressure and temperature, near its critical point, while in the situation analysed here the liquid is localised in the solid, and is highly compressed, owing to the large volume change on melting \cite{Averback}. For the present analysis, one must also note that in the fluid phase of noble solid gases, in particular in Xe, the electronic and atomic subsystems are no more coupled \cite{brown}.

A dynamic theory must take also into account the compressibility and viscosity of the liquid, the elasticity of the solid, heat conduction effects, as well as the influence of surface tension. The subsequent evolution of the spike, after melting, must be characterised using the techniques of fluid dynamics, i.e. the conservation laws of mass, energy and momentum, as well as the thermal and caloric equations of state \cite{Apfel}.

\begin{figure}[!htb]
\centering
\includegraphics[width=0.6\textwidth]{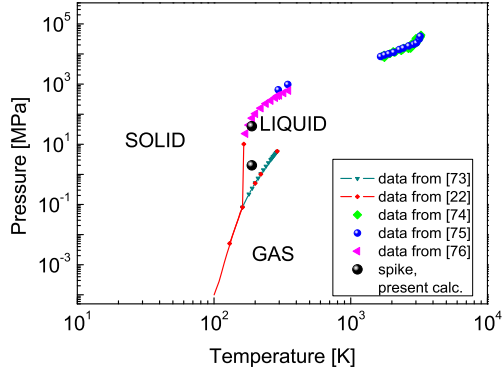}
\begin{small}
\caption{Phase diagram of Xe, and the point corresponding to the maximum temperature and pressure produced by the spike.}
\label{fig9}
\end{small}
\end{figure}

The phase diagram of Xe is presented in Figure 9, using data from \cite{Yoo, Skripov, Boehler, Ross, Wang}.

As a first approximation, we calculated the temperature in the liquid from the energy balance, keeping into account the energy available in the atomic system after the phase transformation, and the pressure induced in the fluid due to the phase transition, supposing the solid incompressible, and consequently the fluid evolving at constant volume. The Stobridge equation of state of liquid xenon obtained in Ref. \cite{Streett} was used:

\begin{equation}
\begin{split}
 p = &RT\rho  + (A_1 RT + A_2  + A_3 /T + A_4 /T^2  + A_5 /T^4 )\rho ^2  + \left( {A_6 RT + A_7 } \right)\rho ^3 + \\
   &+ A_8 T\rho ^4  + \left( {A_9 /T^2  + A_{10} /T^3  + A_{11} /T^4 } \right)\exp \left( {A_{16} \rho ^2 } \right)\rho ^3  +  \\
  &+ \left( {A_{12} /T^2  + A_{13} /T^3  + A_{14} /T^4 } \right)\exp \left( {A_{16} \rho ^2 } \right)\rho ^5  + A_{15} \rho ^6  \\
 \end{split}
 \end{equation}
where $A_1$=$-$1.1255 cm$^3$ g$^{-1}$, $A_2$=40.9801 MPa cm$^6$ g$^{-2}$, $A_3=-$2.6005 $\times 10^4$ MPa cm$^6$ K g$^{-2}$, $A_4=$2.6254 $\times 10^6$ MPa cm$^6$ K$^2$ g$^{-2}$, $A_5=-$1.3782 $\times 10^{10}$ MPa cm$^6$ K$^4$ g$^{-2}$, $A_6=-$0.4521 cm$^6$ g$^{-2}$, $A_7=$10.9393 MPa cm$^9$ K g$^{-3}$, $A_8=$0.0065 MPa cm$^{12}$ K$^{-1}$ g$^{-4}$, $A_9=-$5.2197 $\times 10^5$ MPa cm$^9$ K$^2$ g$^{-3}$, $A_{10}=$5.5818 $\times 10^8$ MPa cm$^9$ K$^3$ g$^{-3}$, $A_{11}=-$6.1875 $\times 10^{10}$ MPa cm$^9$ K$^4$ g$^{-3}$, $A_{12}=-$1.1191 $\times 10^5$ MPa cm$^{15}$ K$^2$ g$^{-5}$, $A_{13}=-$3.8587 $\times 10^7$ MPa cm$^{15}$ K$^3$ g$^{-5}$, $A_{14}=$6.8595 $\times 10^9$ MPa cm$^{15}$ K$^4$ g$^{-5}$, $A_{15}=$0.1594 MPa cm$^{18}$ g$^{-6}$, $A_{16}=-$0.2296 cm$^6$ g$^{-2}$, $R$=0.0630 MPa cm$^3$ K$^{-1}$ g$^{-1}$, and $\rho$ is the density.

The results of the calculation for the time and distance of the maximum atomic temperature produced by a 25 keV Xe selfrecoil are presented in Figure 10. The melting temperature at normal pressure is attained and surpassed. Inside the solid a metastable, compressed liquid exists, localised both in time and space. The value for the pressure attained in the liquid, calculated in this approximation, is around 40 MPa. In the same time, disregarding the elastic properties of the solid, the pressure value inside the liquid, obtained from computations is overestimated. The point corresponding to the maximum temperature and pressure developed in the liquid during the spike is also represented in Figure 9. In a second step, using the compressibility of the liquid and the shear modulus of the solid, at the pressure calculated previously, a volume expansion with around 12 \% is obtained, which in turn leads to a decrease of the pressure up to approximately 2 MPa - also figured.  A more realistic model describing the dynamics of the solid-liquid transition, with the consideration of the elastic properties of the solid and of the shockwave produced, and characterising the liquid phase using fluid dynamics equations, will be developed in a future paper.

\begin{figure}[!htb]
\centering
\includegraphics[width=0.6\textwidth]{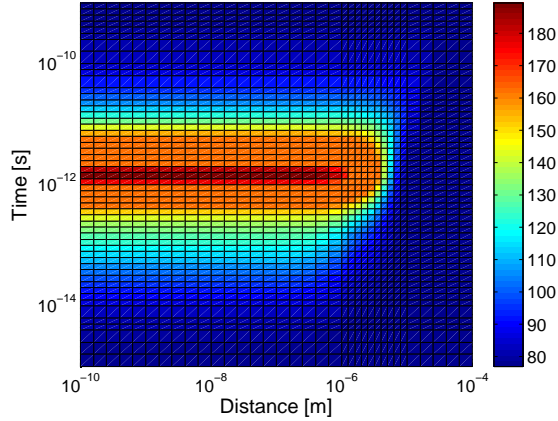}
\begin{small}
\caption{Color map of the atomic temperature produced by a recoil of 25 keV in the Xe crystal kept at 77 K.}
\label{fig10}
\end{small}
\end{figure}

The influence of the recoil energy on the time dependence of the atomic temperature, at fixed distance in respect to the trajectory, was also studied, and is presented in Figure 11 for Xe primary self recoils, at 100 nm from the trajectory, in the material of initial temperature 77 K. As could be seen, for the lowest recoil energy considered in this analysis (3 keV), the melting temperature is not reached. Increasing the kinetic energy of the recoil, the atomic temperature reaches the melting point (starting from 5 keV), a small part of the material remains longer at the melting temperature for the 15 and 20 keV recoils, and this temperature is even surpassed for recoils of kinetic energy higher than 20 keV. For the 25 keV kinetic energy Xe recoil, the maximum temperature reached is 189 K.

\begin{figure}[!htb]
\centering
\includegraphics[width=0.6\textwidth]{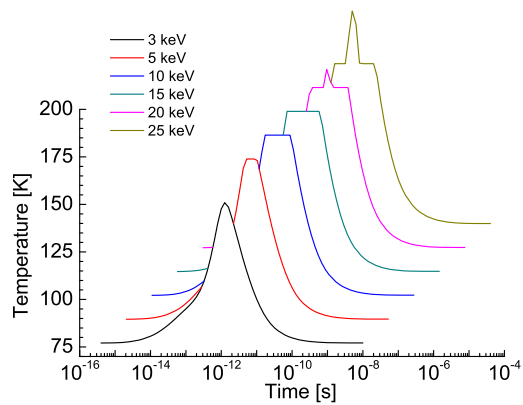}
\begin{small}
\caption{Time development of the heated region at 100 nm from the trajectory of the recoil, for recoils of different energies in Xe at 77 K, putting into evidence the phase transition(s).}
\label{fig11}
\end{small}
\end{figure}

The initial temperature influences the development of the spike in two ways: by the temperature dependence of the physical characteristics of the material, and by the departure in respect to the melting temperature at normal pressure. In Figure 12, the atomic temperature produced by a 15 keV recoil in Ar is represented as a function on the time after the passage of the recoil, for initial temperatures in the range 2 $-$ 70 K. Starting from the initial temperature of 30 K, the melting point is reached. In the figure the horizontal dotted line marks the temperature corresponding to the phase transitions solid - liquid.

\begin{figure}[!htb]
\centering
\includegraphics[width=0.6\textwidth]{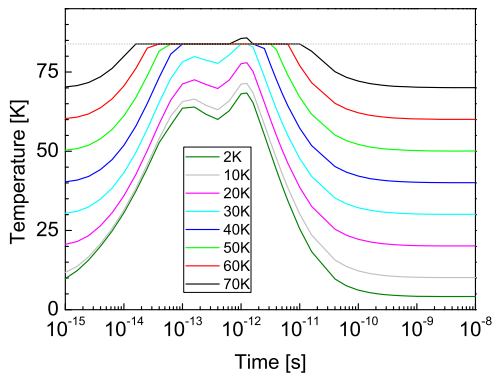}
\begin{small}
\caption{Time development of the thermal pulse in the atomic subsystem of solid Ar, produced by an Ar recoil of 15 keV, for different initial temperatures.}
\label{fig12}
\end{small}
\end{figure}

In Figure 13, the space and time dependencies of the atomic temperature in Ar, Kr and Xe are presented comparatively with the same dependencies for Si and Ge, for the recoils produced by the elastic scattering of a WIMP of 30 and 120 GeV/c$^2$ kinetic energy respectively.

\begin{figure}
\centering
\subfigure{
\includegraphics[width=0.4\textwidth]{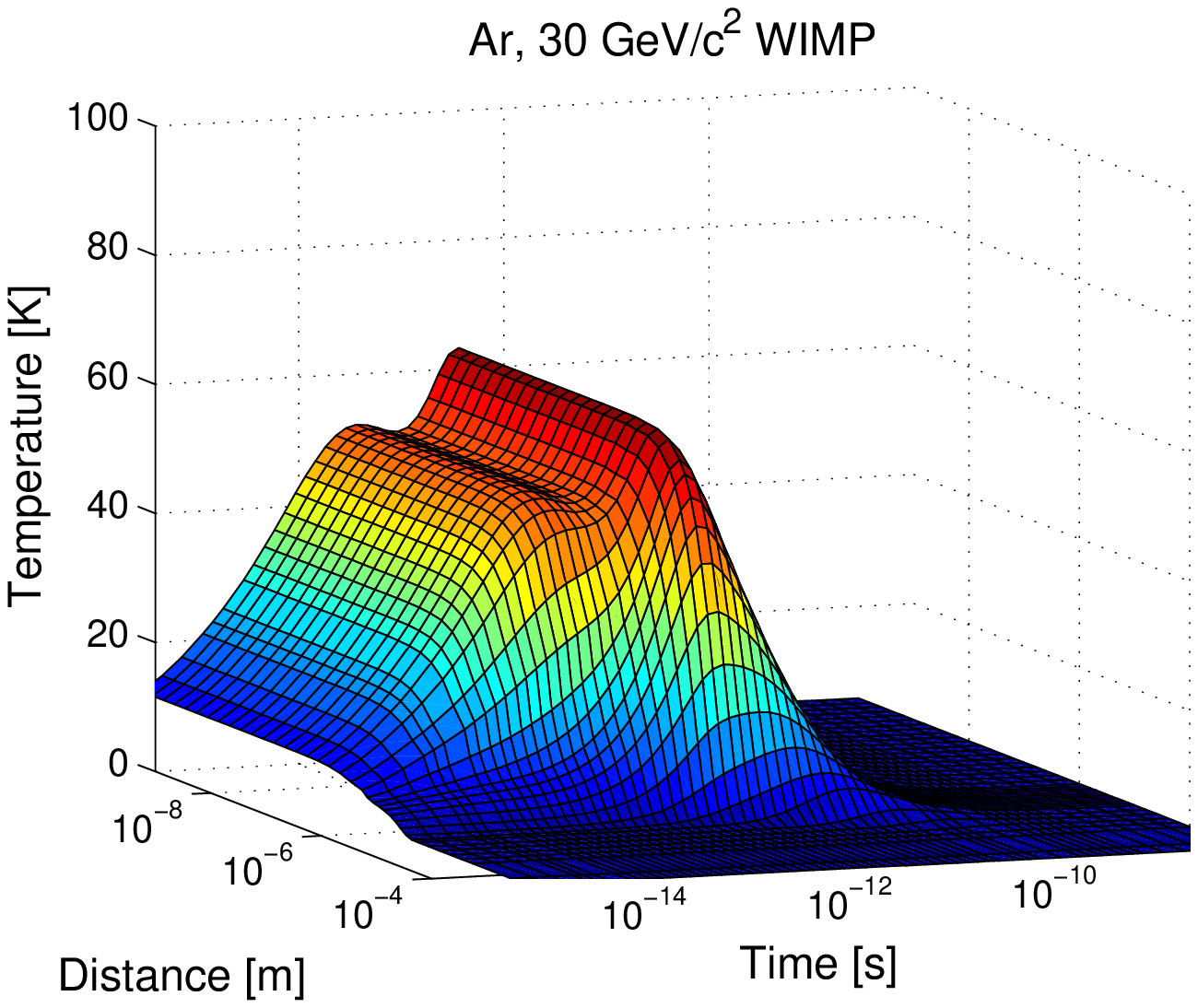}
\label{fig:subfig13a}
}
\subfigure{
\includegraphics[width=0.4\textwidth]{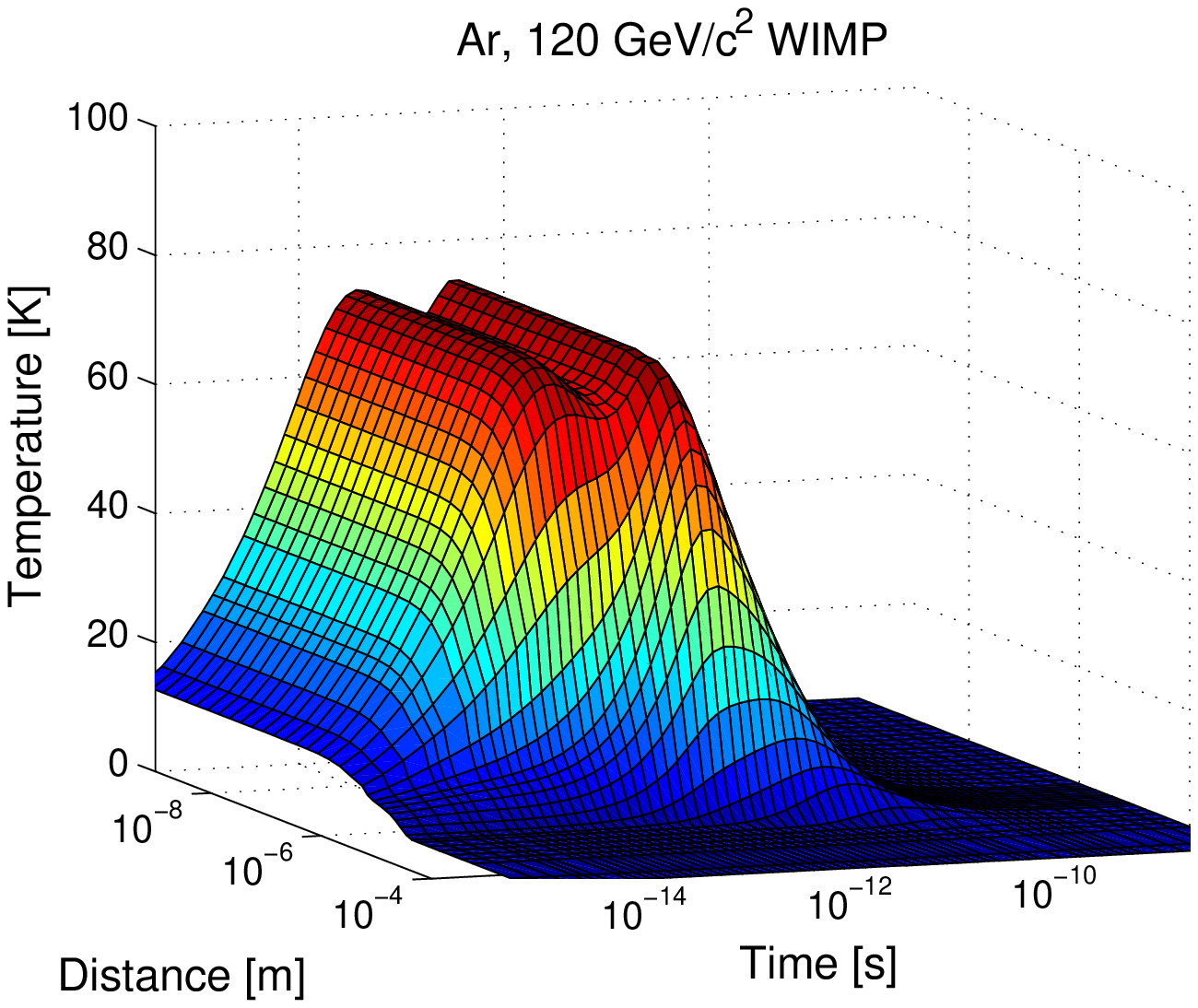}
\label{fig:subfig13b}
}
\subfigure{
\includegraphics[width=0.4\textwidth]{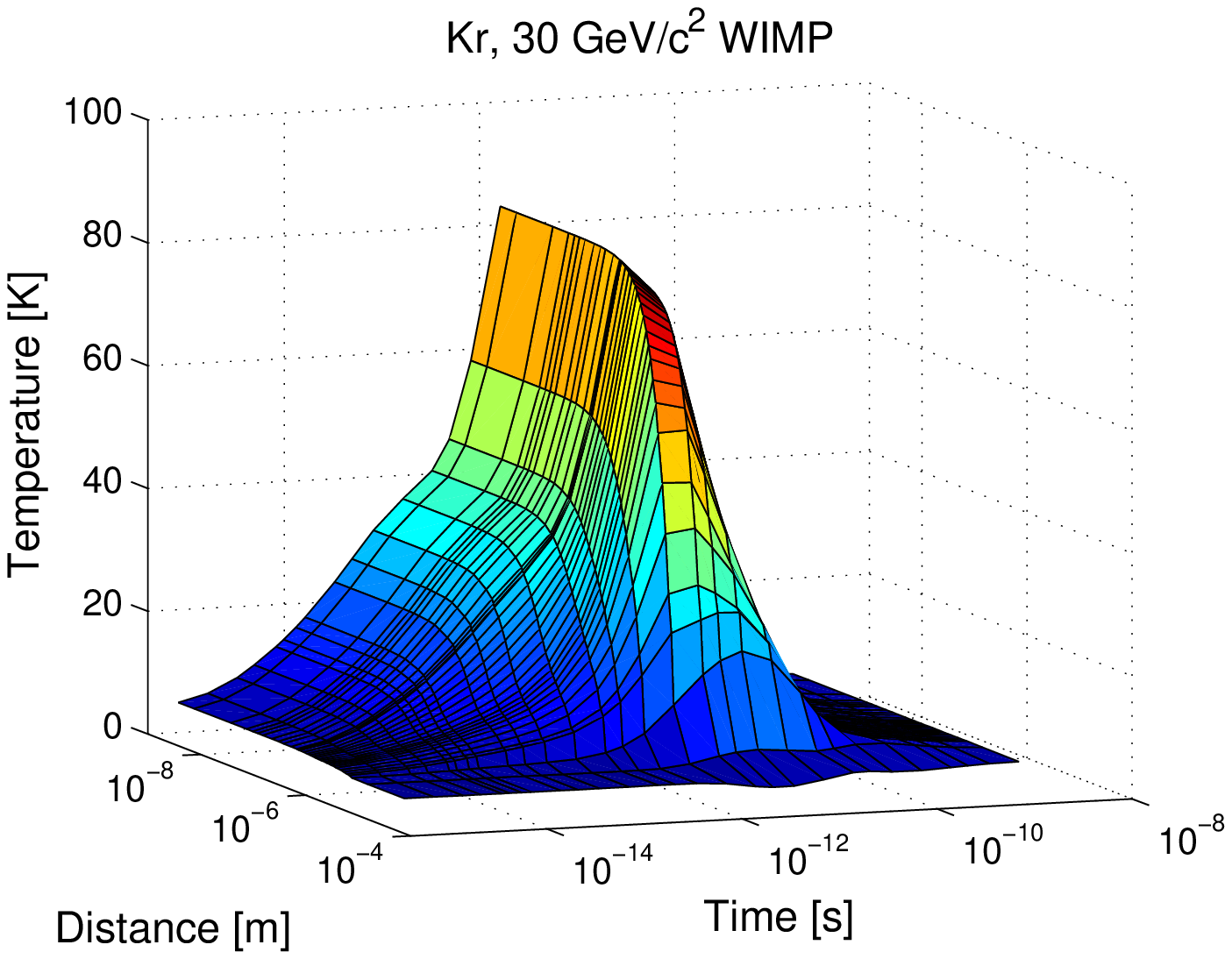}
\label{fig:subfig13c}
}
\subfigure{
\includegraphics[width=0.4\textwidth]{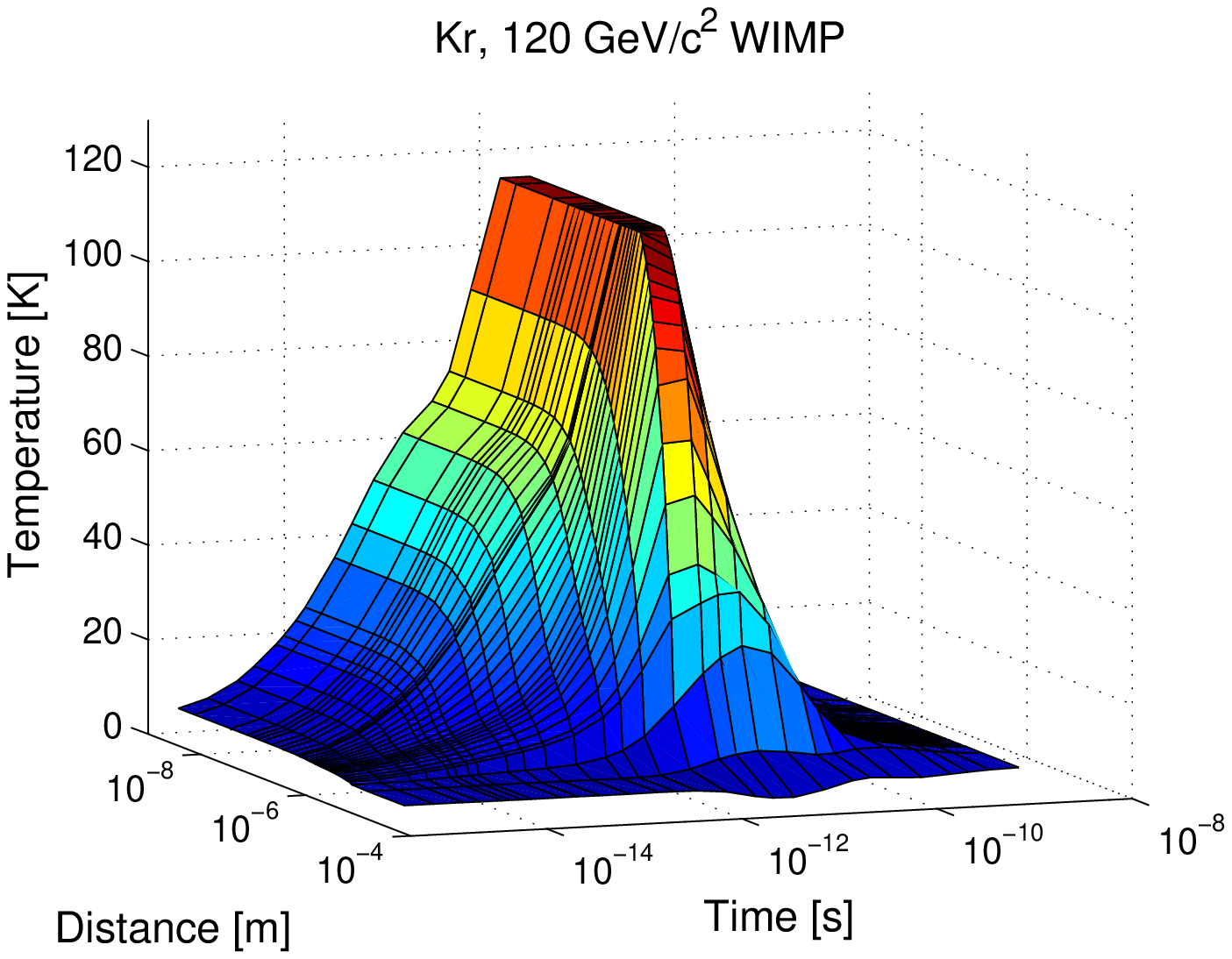}
\label{fig:subfig13d}
}
\subfigure{
\includegraphics[width=0.4\textwidth]{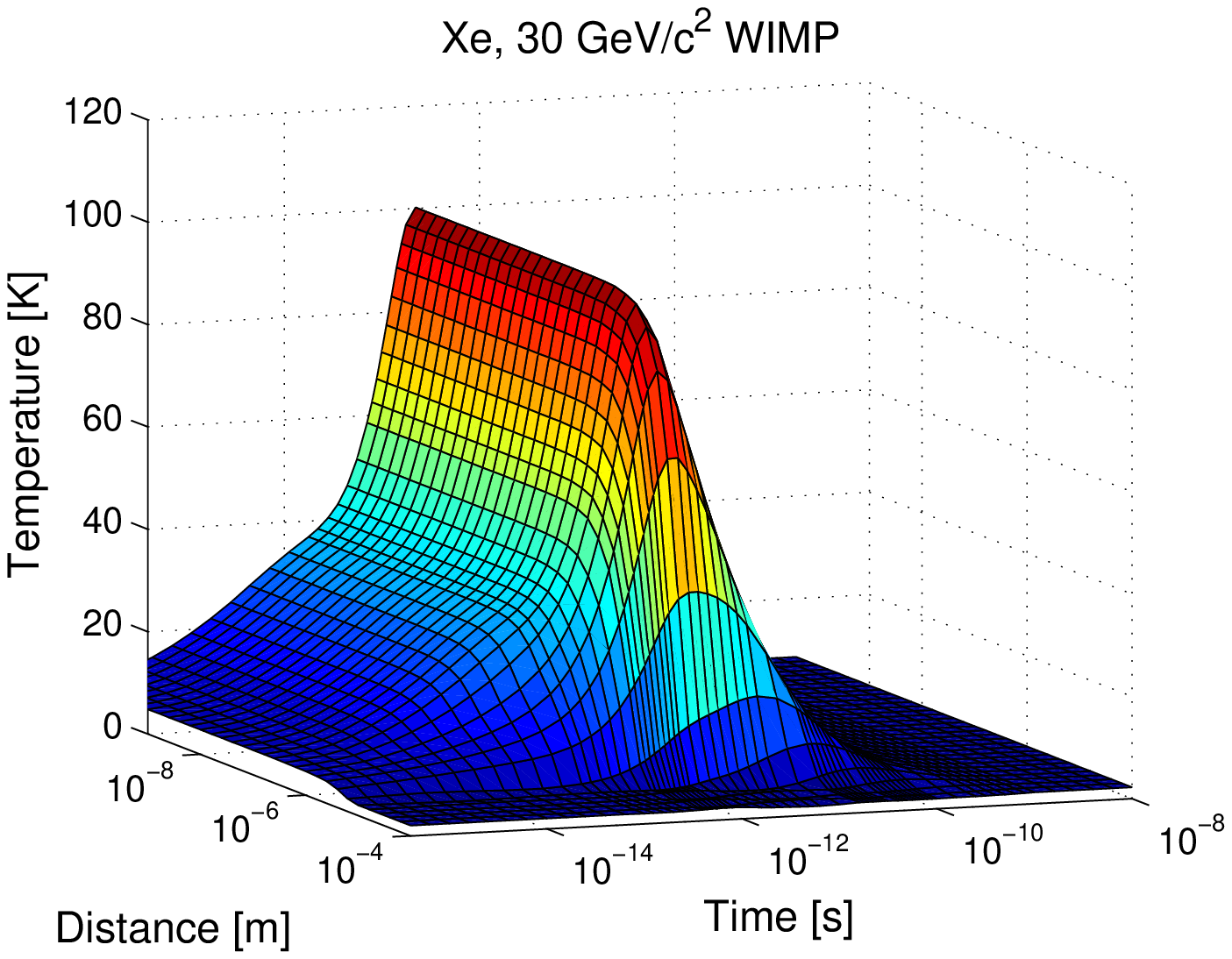}
\label{fig:subfig13e}
}
\subfigure{
\includegraphics[width=0.4\textwidth]{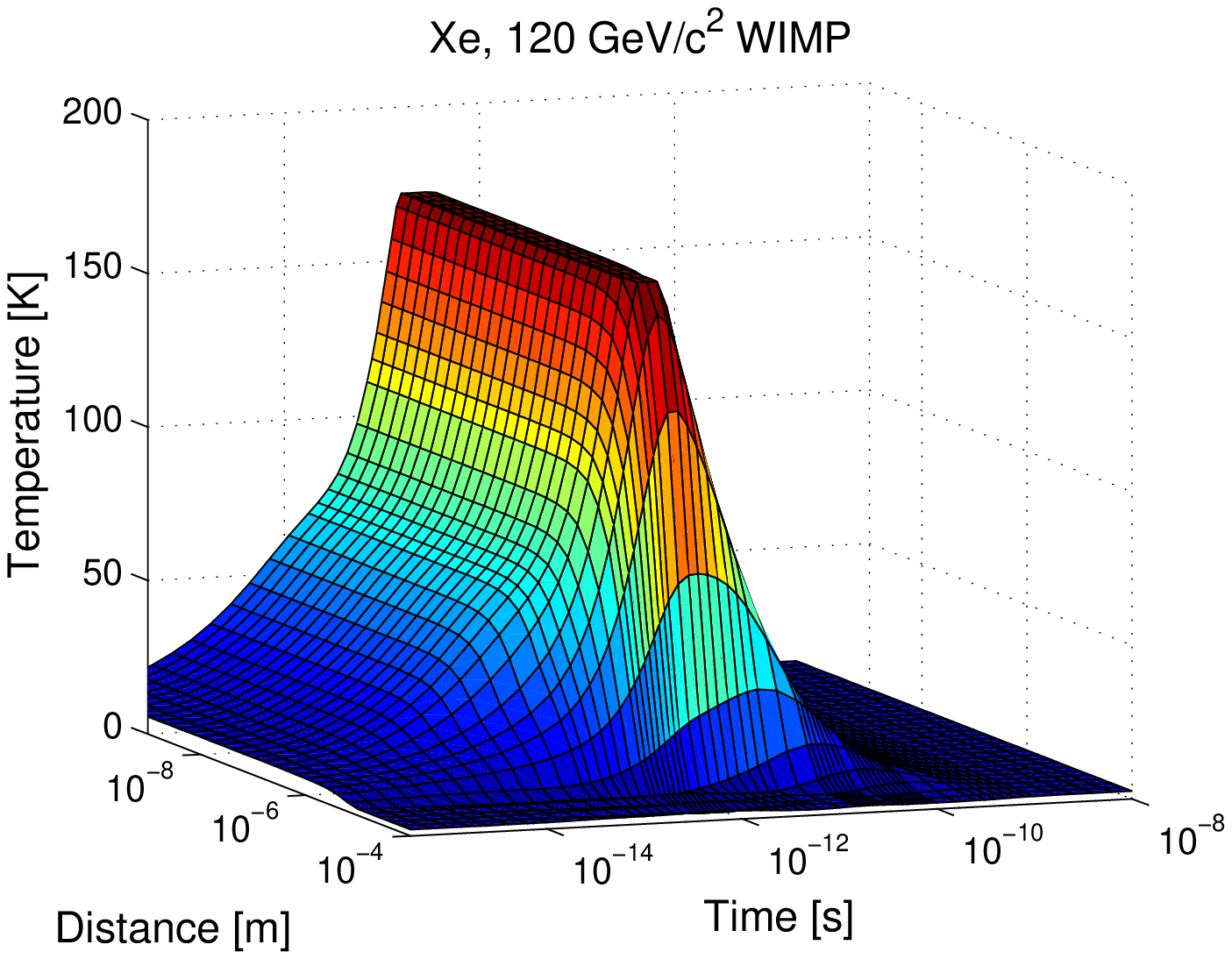}
\label{fig:subfig13f}
}
\subfigure{
\includegraphics[width=0.4\textwidth]{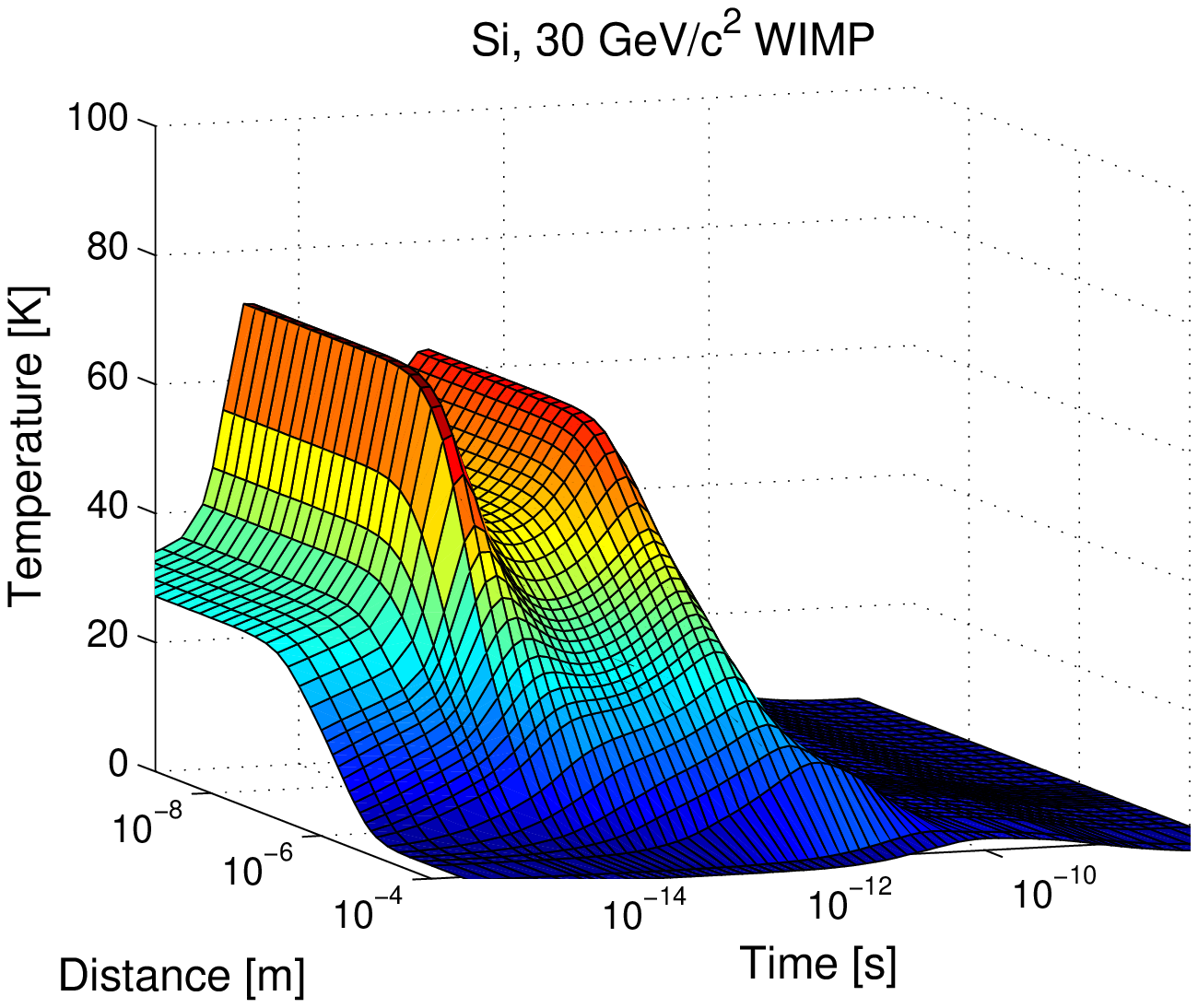}
\label{fig:subfig13g}
}
\subfigure{
\includegraphics[width=0.4\textwidth]{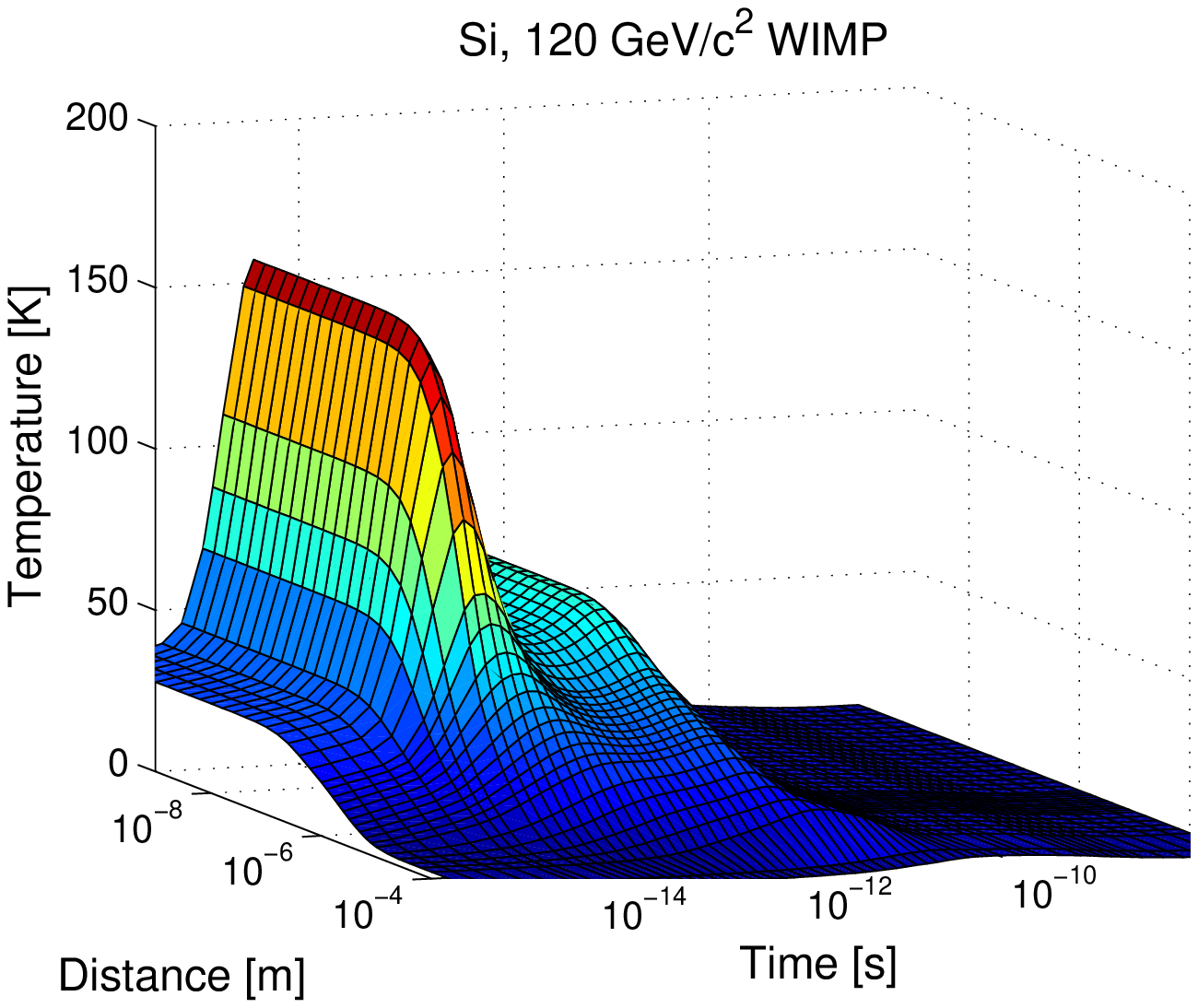}
\label{fig:subfig13h}
}
\subfigure{
\includegraphics[width=0.4\textwidth]{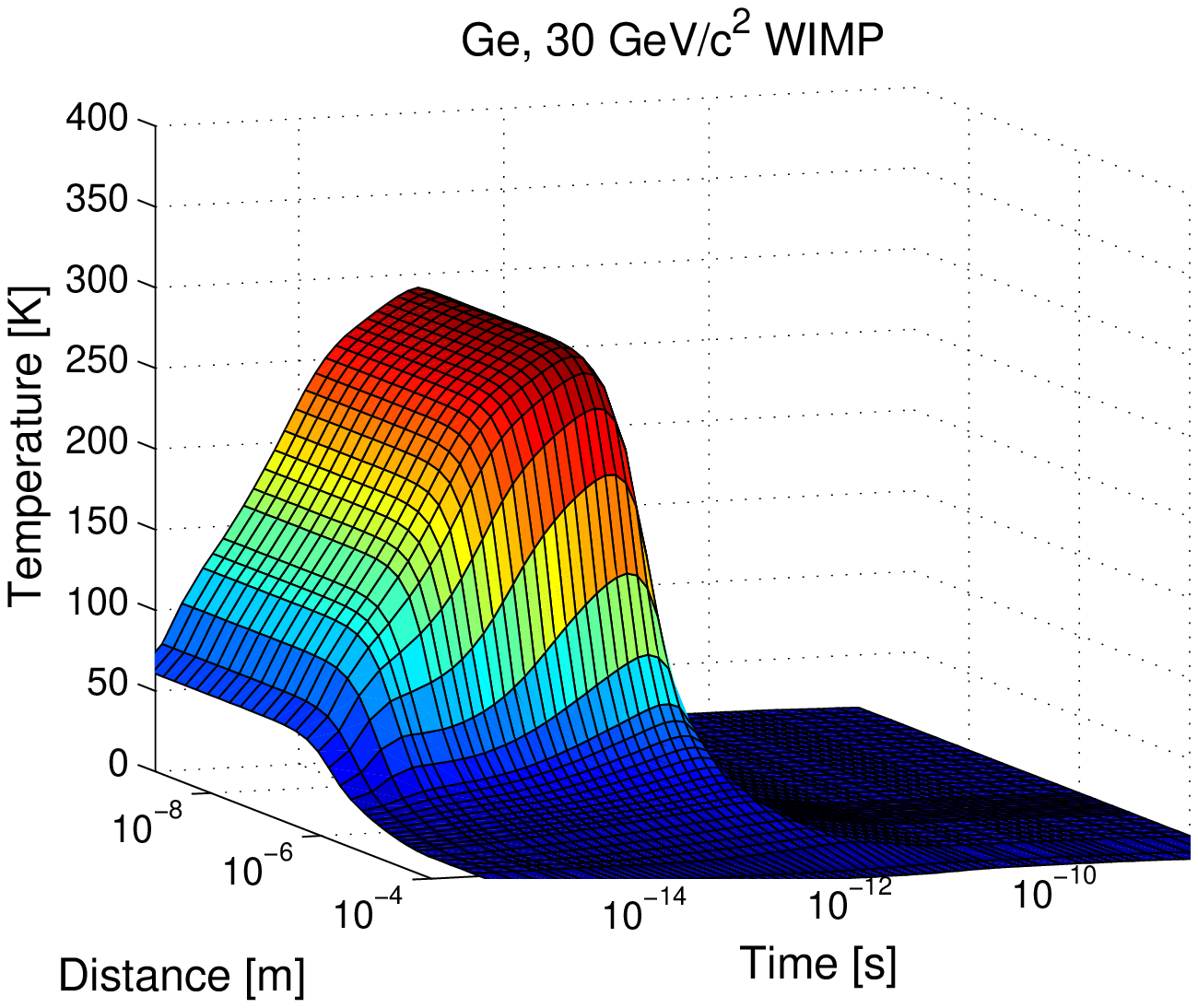}
\label{fig:subfig13i}
}
\subfigure{
\includegraphics[width=0.4\textwidth]{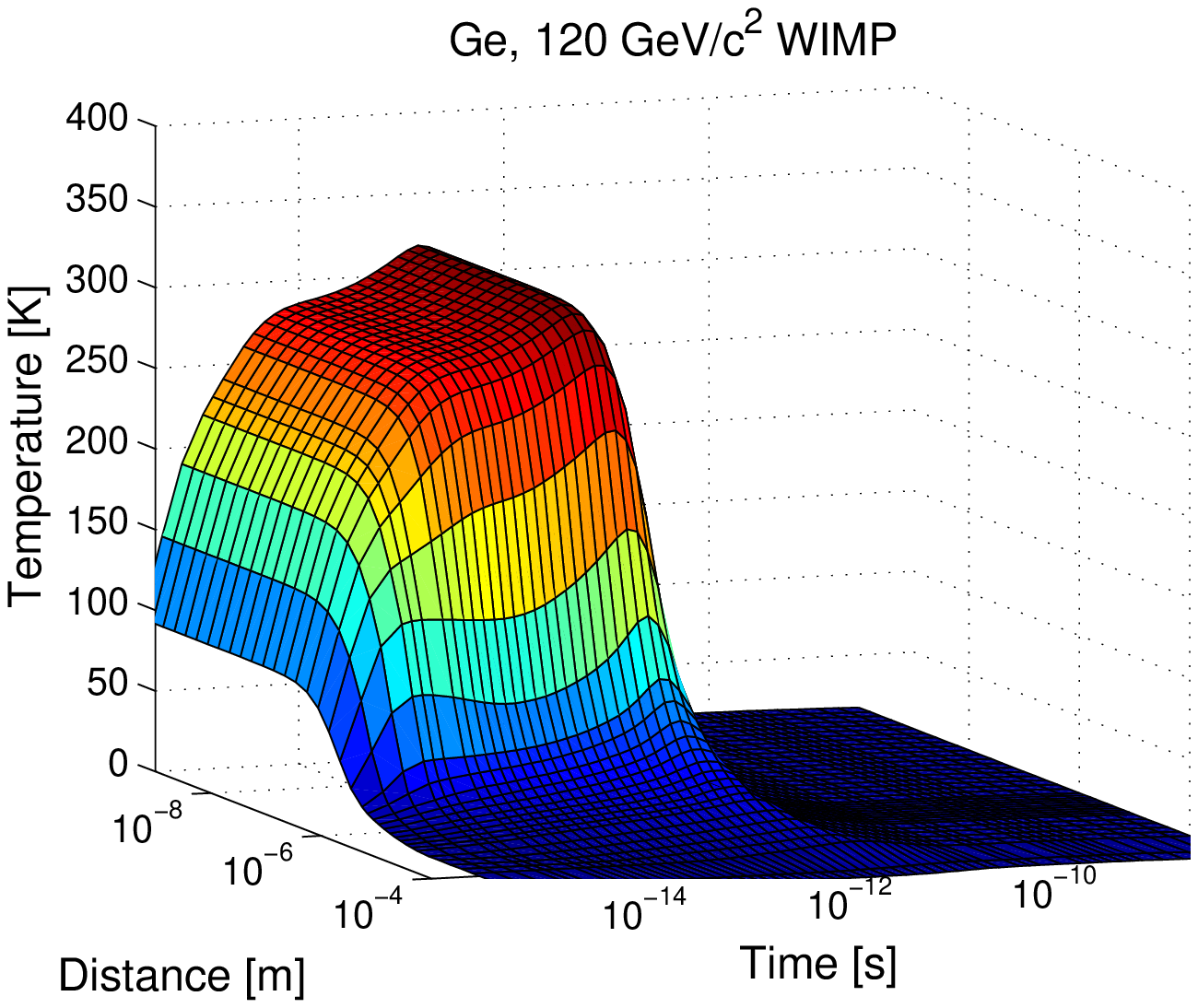}
\label{fig:subfig13j}
}

\begin{small}
\caption{Lattice temperature dependence on distance and time in solid Ar, Kr, Xe, in Si and Ge, as due to a self-recoil produced in the elastic scattering of a WIMP of: 30 GeV/c$^2$ (left) and 120 GeV/c$^2$ (right).}
\label{fig13}
\end{small}
\end{figure}

We would like to emphasize that in Si and Ge at cryogenic temperatures the linear energy transfer between the electron and lattice subsystems is replaced by a term $g\left( {T_e^p  - T_a^p } \right)$, with $p= 5 - 6$. On the other side, the ratio between the electronic and nuclear energy loss decreases from Si, to Ar, Ge, Kr and Xe, and increases slowly with the WIMP mass.
In Ar, the temperature response of the material to the WIMP has a fast and a slow component, similarly to semiconductors, in the whole range of masses investigated. In Kr and Xe the major contribution to the transient in the atomic temperature comes from the nuclear energy loss, and the shoulder in the time dependence of $T_a$ becomes less visible for Xe. The phase transition is produced for Kr and Xe, for the 120 GeV/c$^2$ WIMP.

In Figure 14, the correlated parameters (solid gas, recoil energy) corresponding to the threshold energy of the recoil which produces the solid - liquid phase transition are represented as a function of the temperature of the material for Ar, Kr and Xe. The band corresponds to the imprecision in the recoil energy for a defined temperature of the sample, in the numerical model calculations. It is important to mention that in the case of Ar the phase transition becomes possible only at higher temperatures, due to the fact that the maximum of the nuclear energy loss corresponds to around 25 KeV.
For an established temperature of the medium, the phase transition might be experimentally used as a supplementary information about the energy of the recoil.

\begin{figure}[!htb]
\centering
\includegraphics[width=0.6\textwidth]{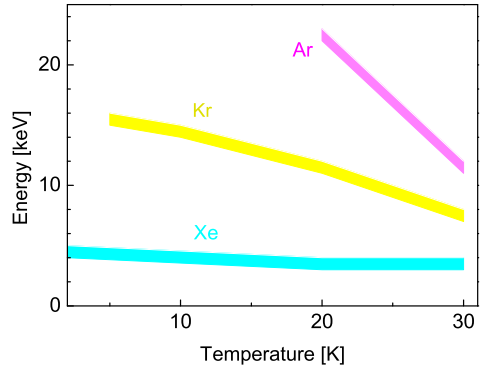}
\begin{small}
\caption{Threshold recoil energy for the solid - liquid phase transition in solid Ar, Kr and Xe as a function of the temeprature of the samples.}
\label{fig14}
\end{small}
\end{figure}

\section{Conclusions}
The transient phenomena produced in solid noble gases by the stopping of the recoils resulting from the elastic scattering processes of WIMPs from the galactic halo were modelled, as dependencies of the temperatures of lattice and electronic subsystems on the distance to the recoil's trajectory, and time from its passage.

The peculiarities of these thermal transients produced in Ar, Kr and Xe were analysed for different initial temperatures and WIMP energies, and were correlated with the characteristics of the targets and with the energy loss of the recoils. The results were compared with the thermal spikes produced by the same WIMPs in Si and Ge.

In the range of the energy of interest, up to tens of keV for the self-recoil, the local phase transition solid - liquid was found possible, and the threshold parameters were established.

\section*{Acknowledgements}
We are grateful to Prof. Gh. Ciobanu for useful discussions regarding the phenomena accompanying phase transitions. S.L. would like to thank UEFISCDI for support, under Project PNII - IDEI, 901/2008.

\end{document}